\documentclass[conference]{IEEEtran}
\IEEEoverridecommandlockouts
\usepackage{cite}
\usepackage{amsmath,amssymb,amsfonts}
\usepackage{algorithmic}
\usepackage{graphicx}
\usepackage{textcomp}
\usepackage{enumitem}
\usepackage{xcolor}
\usepackage{subcaption}
\usepackage{placeins}
\usepackage[nolist]{acronym}
\usepackage{float}
\usepackage{tabularx}
\usepackage{xspace}
\usepackage[export]{adjustbox}
\usepackage{hyperref}

\begin{acronym}
    \acro{PU}{Processing Unit}
    \acrodefplural{PU}[PUs]{Processing Units}
    \acrodefplural{DIMM}[DIMMs]{Dual In-line Memory Module}
    \acro{DIMM}{Dual In-line Memory Module}
    \acro{MC}{Memory Controller}
    \acro{ACT}{Activate}
    \acro{PRE}{Pre-charge}
    \acro{CAS}{Column Address Strobe}
    \acro{REFab}[REF\textsubscript{ab}]{All-bank Refresh}
    \acro{REFsb}[REF\textsubscript{sb}]{Same-bank Refresh}
    \acro{WCRT}{Worst Case Response Time}
    \acro{WCET}{Worst Case Execution Time}
    \acro{PRAC}{Per-Row-Activation-Counter}
    \acro{ABO}{Alert-Back-Off}
    \acro{RFM}{Refresh Management}
\end{acronym}

\def\BibTeX{{\rm B\kern-.05em{\sc i\kern-.025em b}\kern-.08em
    T\kern-.1667em\lower.7ex\hbox{E}\kern-.125emX}}

\newcommand{\floor}[1]{\lfloor #1 \rfloor}
\newcommand{\ceil}[1]{\lceil #1 \rceil}

\newcommand{\CR}[1]{{#1}} 
\newcommand{\TR}[1]{{}} 
\newcommand{\Jenga}{\textsc{Jenga}\xspace}

\usepackage[most]{tcolorbox}
\newtcolorbox{researchquestion}[1][]{
    enhanced,
    colback=blue!5!white,
    colframe=blue!65!black,
    boxrule=1pt, 
    top=3pt,                    
    bottom=3pt,
    left=4pt,
    right=4pt,
    #1
}
\begin{document}

\title{\Jenga: Exploiting Counter-Based RowHammer Countermeasures to Break Real-Time Predictability
}

\author{
\IEEEauthorblockN{
Valentin Abgrall\textsuperscript{1},
Marcello Traiola\textsuperscript{1},
Ruben Salvador\textsuperscript{1,2},
Alessandro Palumbo\textsuperscript{1,2}\\
Maria Mendez Real\textsuperscript{3},
Angeliki Kritikakou\textsuperscript{1}
}

\IEEEauthorblockA{\textsuperscript{1}Inria Rennes / IRISA Lab, Univ Rennes}
\IEEEauthorblockA{\textsuperscript{2}CentraleSupélec}
\IEEEauthorblockA{\textsuperscript{3}Lab-STICC, Univ Bretagne Sud}
}

\author{
\IEEEauthorblockN{Valentin Abgrall}
\IEEEauthorblockA{
Univ Rennes, CNRS, Inria \\
IRISA - UMR 6074, F-35000 Rennes \\
valentin.abgrall@irisa.fr
}
\\
\IEEEauthorblockN{Maria Méndez Real}
\IEEEauthorblockA{
Univ. Bretagne-Sud \\
Lab-STICC UMR CNRS 6285 \\
maria.mendez-real@univ-ubs.fr
}
\and
\IEEEauthorblockN{Marcello Traiola}
\IEEEauthorblockA{
Univ Rennes, CNRS, Inria \\
IRISA - UMR 6074, F-35000 Rennes \\
marcello.traiola@inria.fr
}
\\
\IEEEauthorblockN{Alessandro Palumbo}
\IEEEauthorblockA{
CentraleSupélec, CNRS, Inria \\
IRISA - UMR 6074, F-35000 Rennes \\
alessandro.palumbo@inria.fr
}
\and
\IEEEauthorblockN{Ruben Salvador}
\IEEEauthorblockA{
CentraleSupélec, CNRS, Inria \\
IRISA - UMR 6074, F-35000 Rennes \\
ruben.salvador@inria.fr
}
\\
\IEEEauthorblockN{Angeliki Kritikakou}
\IEEEauthorblockA{
Univ Rennes, CNRS, Inria \\
IRISA - UMR 6074, F-35000 Rennes \\
angeliki.kritikakou@irisa.fr
}
}

\maketitle

\begin{abstract}
Safety-critical real-time systems must satisfy multiple dependability requirements, notably time predictability and security. In such systems, tasks must complete within bounded and known execution times, typically characterised through Worst-Case Execution Time (WCET) analysis. At the same time, DRAM-based platforms are increasingly sensitive to the RowHammer read-disturbance security vulnerability, which has motivated the development of numerous hardware and software countermeasures in both academia and industry. However, the impact of these defences is generally evaluated in terms of average-case performance, a metric that is insufficient for safety-critical real-time systems, where worst-case behaviour is the primary concern.

In this paper, we study the impact of RowHammer countermeasures based on hardware counters on the timing behaviour of real-time systems. We use a Per-Row-Activation-Counter (PRAC) countermeasure as a case study, \CR{standardised for} recent DDR5 memories, and show that it can introduce significant timing variations. Based on this observation, we introduce \Jenga, an attack in which an attacker-controlled task manipulates the internal state of the RowHammer countermeasure mechanism to increase the execution time of a victim real-time task beyond its expected WCET. We implement \Jenga in a gem5 and Ramulator 2.0 simulation environment and evaluate its impact on TACLeBench workloads. We show that such an attack can delay tasks up to 200\% of their WCET, making the initial time-safety assumptions unsafe. To address this issue, we derive a safe analytical bound that accounts for mitigation-induced delays in WCET analysis for DRAM systems protected by hardware countermeasures, such as PRAC-N.
\end{abstract}


\section{Introduction}


Safety-critical real-time systems, such as avionics, Unmanned Aerial Vehicles (UAVs), autonomous vehicles, and medical devices, must satisfy stringent dependability requirements. In particular, tasks executing on these systems must complete within bounded and predictable execution times, as timing violations may lead to catastrophic consequences, including mission failure or threats to human safety \cite{rts_meet_deadline}. Historically, the security of such systems has often received less attention than their timing and functional correctness properties, partly due to the assumption that embedded real-time platforms were unlikely targets for attackers \cite{sok_rts}. However, this assumption has been challenged over the last decade by numerous demonstrations of attacks targeting real-time and cyber-physical systems, including connected vehicles and UAVs \cite{jeep,drones,dronesec}. Consequently, modern critical systems must now simultaneously ensure both strong timing guarantees and robust security properties.

A major challenge is that security and real-time analysis have traditionally been investigated by distinct research communities. In practice, security mechanisms are not timing-neutral: they can introduce additional computations, memory accesses, resource contention, etc., thereby increasing execution-time overheads and timing variability \cite{overhead_security_2}. While such overheads are generally acceptable in general-purpose systems, they can become problematic in safety-critical real-time systems, where even moderate increases in execution time may compromise schedulability or invalidate previously established \ac{WCET} guarantees \cite{sok_rts}. To address this limitation, on the one hand, security mechanisms should be designed with timing predictability in mind in order to be integrated into real-time systems. On the other hand, real-time approaches should account for the impact of security mechanisms on execution during WCET estimation and real-time guarantees. 

Among well-known cybersecurity threats, the RowHammer vulnerability has emerged as a major concern for DRAM-based systems since its first disclosure in 2014\cite{rowhammer_first_paper}. Due to physical coupling effects between adjacent DRAM cells, repeatedly activating (i.e., “hammering”) a given aggressor row can induce bit flips in neighbouring victim rows. In practice, RowHammer can be exploited at the software level to perform privilege escalation \cite{double, mobile_rh} or to bypass logical memory isolation \cite{rh_vm, rh_js}. Since 2014, sophisticated hammering patterns have been proposed \cite{double, half, one_loc_rh, sledgehammer, blacksmith, marionette}, progressively improving the attack's effectiveness and stealthiness. This vulnerability is worsening with new DRAM generations. As DRAM feature sizes decrease, cell capacitance is reduced, and cells become increasingly susceptible to charge leakage and inter-cell interference~\cite{revisiting_rh}. Consequently, RowHammer represents a critical reliability and security concern for contemporary and future DRAM-based systems.
Defending against RowHammer is very challenging, as it results from a hardware vulnerability that can be triggered through purely software-controlled memory accesses. Multiple software-level mitigations proposed in the literature have later been shown to be bypassable~\cite{one_loc_rh}. As a result, recent research has increasingly focused on hardware-level mitigations integrated directly into DRAM devices or memory controllers~\cite{survey_rh_mitig}. The majority of the proposed solutions follow a common paradigm: they monitor activations to different physical memory regions through counters and trigger a preventive mitigation action (usually refreshing a set of rows) when a specific condition on the counters is met.

These hardware countermeasures introduce overheads that remain reasonable when the RowHammer detection threshold is sufficiently high (e.g., when the triggering condition is set to 1000 row activations~\cite {break}). However, as the threshold decreases with technology scaling~\cite{revisiting_rh}, mitigation actions must intervene more frequently, increasing memory interference and degrading overall system performance. This effect has led to the emergence of performance-oriented attacks~\cite{rogue_rfm, not_so_ref, moat}, which deliberately trigger repeated mitigative actions to stall the memory subsystem and disrupt normal operation. To date, the impact of such attacks has been evaluated in terms of average performance degradation in general-purpose systems. However, to the best of our knowledge, their potential impact on real-time systems has not yet been assessed.

This paper addresses this limitation by assessing the time safety of counter-based preventive RowHammer countermeasure in the context of real-time systems. We consider a scenario in which an attacker task accesses the memory within its own address space to alter the state of the hardware counters associated with the RowHammer countermeasure. This modification is referred to as ``building the \Jenga tower". By doing so, the attacker introduces delays in the execution time of the victim task, resulting from cascading mitigations, i.e., the collapse of the \Jenga tower. We use the \ac{PRAC} countermeasure as a case study \cite{jedec_prac}. We show that such an attack can delay tasks up to 200\% of their \ac{WCET}, making the initial time-safety assumptions unsafe.
Building on this finding, we provide a theoretical analysis of the impact of this scenario on task WCET and establish new bounds for tasks executing on DRAM memory protected against RowHammer. 

In summary, this work makes the following contributions:
\begin{itemize}
    \item We develop \Jenga\footnote{\Jenga code is available on the following \href{https://gitlab.inria.fr/vabgrall/jenga}{Gitlab repository}}, a software attack against systems with RowHammer hardware countermeasures based on counters, capable of delaying real-time tasks by up to 200\% of their \ac{WCET}.
    \item We develop a theoretical analysis to compute the \ac{WCET} of programs running on a single-core system protected against RowHammer via a hardware-counter-based countermeasure.
    \item We apply our method to the \ac{PRAC} countermeasure \CR{standardised for} recent DDR5 memories and derive a safe analytical WCET bound.
    \item We run an extensive simulation campaign based on gem5 and Ramulator 2.0 using TACLeBench workloads. The results show that \Jenga effectively interferes with real-time guarantees across all considered workloads. Additionally, we empirically validate our analytical WCET bound for \CR{ten} characteristic workloads.
\end{itemize}


\section{Background}
\label{sec:background}

\subsection{DRAM Organisation and Operations}

\begin{figure}
    \centering
    \includegraphics[width=0.9\linewidth]{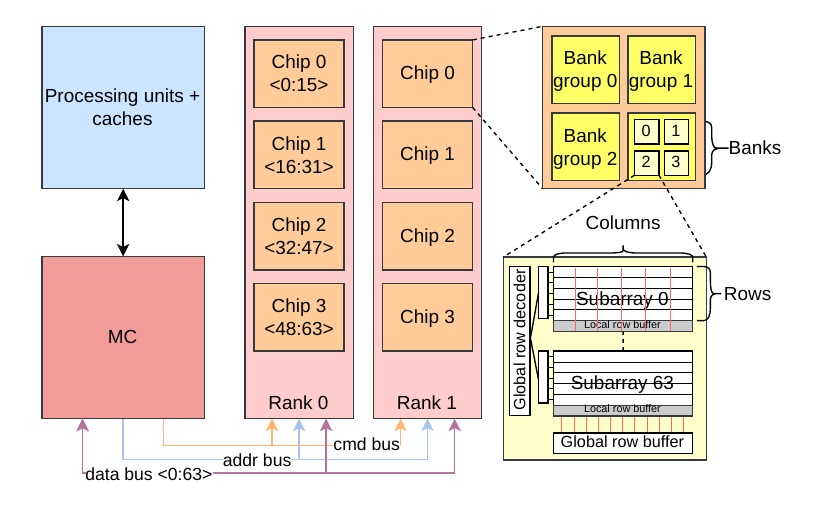}
    \caption{System with one channel and a 2Rx4 DDR4/5 memory module}
    \label{fig:DRAM_org}
\end{figure}

\textbf{Organisation.} In modern systems, DDR4 and DDR5 memories are interfaced with \acp{PU} through a \ac{MC}. The main memory subsystem is hierarchically organised into channels, modules, ranks, chips, bank groups, banks, subarrays, rows, columns and cells, as illustrated in Figure~\ref{fig:DRAM_org}. Each channel operates independently and is connected to its own command, address and data buses. While modules, ranks, chips, bank groups and banks exhibit a certain degree of parallelism, they share common command and data buses within a channel.

\textbf{Addressing.} To access a specific zone in memory, the \acs{PU} issues a 32-bit physical address. The \ac{MC} then translates this address into a tuple $\mathit{(Ch,Md,Rk,Bk,Row,Col)}$, representing the channel, module, rank, bank, row and column containing the requested data. The \textbf{physical-address-to-DRAM-layout} mapping is typically proprietary and not exposed to applications running on the \acs{PU} \cite{knock_knock}. Prior work has shown that this mapping is often linear \cite{rowhammer_first_paper,drama}, as such designs aim to improve memory subsystem performance. In particular, physically contiguous addresses may be placed in distant DRAM regions to exploit parallelism across channels, modules, ranks, and banks.

\textbf{Basic operations.} To access data, the \ac{MC} issues a sequence of DRAM commands over the command bus. A row access begins with an \ac{ACT} command, which activates the target row and transfers its contents into the row buffer of the corresponding bank. Once the row is activated, a \ac{CAS} command is issued to perform a read or write operation on the target column within the row buffer. If a different row in the same bank must be accessed, it creates a \textit{row conflict}: the currently open row must first be closed using a \ac{PRE} command before issuing a new \ac{ACT}. Each command must satisfy strict timing constraints to guarantee correct operation \cite{jedec_prac}. 

\textbf{Access policy.} The \ac{MC} can employ either an open-row or a closed-row policy. Under an open-row policy, an activated row remains open until the \ac{PU} issues a request targeting a different row, at which point a row conflict occurs. Under a closed-row policy, the \ac{MC} keeps a row open only for a predetermined number of requests (its \textit{cap}) before issuing a \ac{PRE} command to close it, even if no other row is accessed in the meantime. An aggressive closed-row policy closes the row immediately after its first access (cap = 1). While open-row policies often provide better overall performance, the row conflicts they introduce can lead to increased timing unpredictability~\cite{WCET_ref}.

\textbf{Refresh operations.} In addition to regular read and write operations, DRAM cells require periodic refresh operations to preserve data integrity due to charge leakage. In DDR5, refreshes can be performed using either \ac{REFab} or \ac{REFsb} commands. The \ac{REFab} command stalls all banks within a rank while some rows are refreshed internally. The DRAM internally manages which rows should be refreshed upon receiving each refresh command. In contrast, \ac{REFsb} only stalls the targeted bank across all bank groups, allowing other banks to continue serving memory requests concurrently. Since the \ac{REFsb} mechanism is not supported in DDR4, we assume in this work that all periodic refreshes are performed using \ac{REFab} commands. We simply call them REF commands.

\subsection{The RowHammer attack}

\begin{figure}
    \centering
    \includegraphics[width=0.8\linewidth]{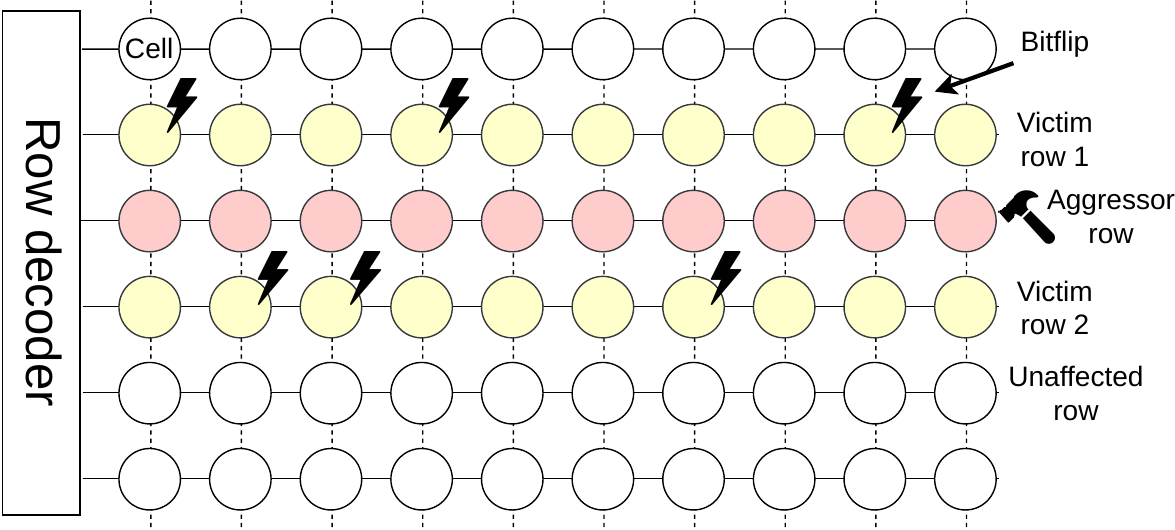}
    \caption{Principle of the RowHammer attack}
    \label{fig:rowhammer}
\end{figure}

RowHammer is a DRAM read-disturbance vulnerability first disclosed in 2014~\cite{rowhammer_first_paper}. Due to physical coupling effects between adjacent DRAM cells, repeatedly activating (i.e., “hammering”) a given \emph{aggressor} row can induce bit flips in neighbouring victim rows, as illustrated in Figure~\ref{fig:rowhammer}. Typically, only a subset of DRAM cells exhibits vulnerability to RowHammer, while others remain stable under identical operating conditions~\cite{rowhammer_first_paper}. For a given DRAM row, there exists a disturbance threshold, denoted $\mathit{T_{RH}}$, corresponding to the maximum number of activations that can be issued within a refresh interval without inducing bit flips. When this threshold is exceeded, one or more bits in adjacent rows may flip.

Since its initial disclosure, RowHammer has evolved significantly. More sophisticated access patterns have been proposed~\cite{double, half, one_loc_rh, sledgehammer, blacksmith, marionette}, progressively improving attack effectiveness, stealthiness, and blast radius (i.e., the distance between aggressors and victims). Beyond RowHammer, additional read-disturbance mechanisms, such as RowPress~\cite{rowpress} and ColumnDisturb~\cite{columndisturb}, have been identified, further worsening the read disturbance issue. In practice, these vulnerabilities can be exploited at the software level to escalate privileges within a UNIX system \cite{double}, to escape a sandboxed environment \cite{double}, to take over an Android system by controlling a single application running with no permission \cite{mobile_rh}, to perform cross-VM (Virtual Machine) attacks \cite{rh_vm} or even to remotely take control of a server by executing JavaScript code \cite{rh_js}.

Finally, technology scaling exacerbates read-disturbance effects. As DRAM feature sizes decrease, cell capacitance is reduced, and cells become increasingly susceptible to charge leakage and inter-cell interference~\cite{revisiting_rh}. Consequently, RowHammer and related disturbance attacks represent a critical reliability and security concern for contemporary and future DRAM-based systems.

\subsection{Rowhammer Countermeasures}\label{sec:rh_mitig}

Defending against RowHammer is challenging because it originates from a hardware vulnerability that can be triggered through purely software-controlled memory accesses. In commodity systems, eliminating the vulnerability entirely would require replacing the underlying DRAM devices. Therefore, for already deployed systems, only software-level countermeasures can be applied. These patches aim at preventing practical exploitation rather than removing the root cause of the vulnerability. For example, ANVIL monitors memory access patterns of running processes and refreshes potential victim rows \cite{anvil}. CATT relies on physical memory isolation \cite{catt}. However, many such countermeasures have later been shown to be bypassable~\cite{one_loc_rh}.

\begin{figure}
    \centering
    \includegraphics[width=0.85\linewidth]{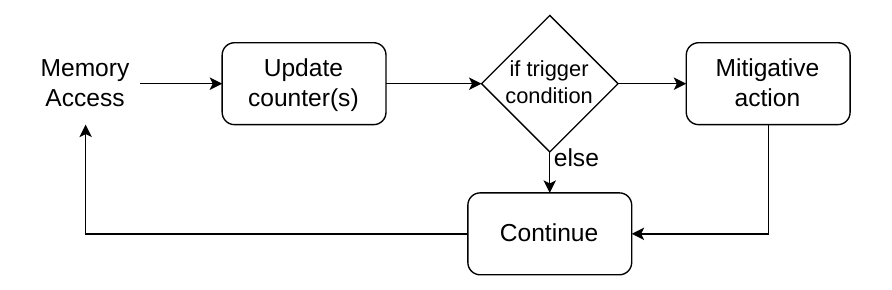}
    \caption{Most RowHammer HW countermeasures follow this paradigm}
    \label{fig:mitig_paradigm}
\end{figure}

As a result, recent research has increasingly focused on hardware-level countermeasures integrated directly within DRAM devices or memory controllers. Most proposed solutions follow a common paradigm, illustrated in Figure~\ref{fig:mitig_paradigm}. They monitor \ac{ACT} commands and maintain activation counters for rows, groups of rows or banks. When a predefined condition is met (for instance, when an activation threshold is exceeded), a mitigation action is triggered (typically an additional refresh of potential victim rows~\cite{survey_rh_mitig}). The main differences between existing mechanisms lie in how activations are tracked and in the spatial granularity at which vulnerable rows can be identified. We use the PRAC countermeasure, \CR{standardised for} commercial DRAM devices, as a case study.

\textbf{\ac{PRAC}}: introduced in the DDR5 JEDEC standard \cite{jedec_prac}, \ac{PRAC}-N associates a \textbf{hardware} counter with each DRAM row. \CR{The counters are internal to the DRAM and are not software-accessible.} As per the standard, a counter is incremented whenever its row is activated, either by an \ac{ACT}, REF, or \ac{RFM} command \cite{pvac}. When a manufacturer-defined condition on the counters is satisfied, the DRAM issues an \ac{ABO} alert to the \ac{MC}. Upon receiving this command, the \ac{MC} continues normal operation for 180ns before issuing $N$ (1, 2, or 4) back-to-back \ac{RFM} commands to the DRAM in order to refresh the corresponding victim rows. Each of these commands refreshes the neighbours of $\mathit{n_{RFM}}$ aggressor rows, since they may experience charge leakage due to their proximity to the aggressor rows; eventually, they could also be flagged as aggressors if their counters exceed the detection threshold after the refresh.

\begin{figure}
    \centering
    \includegraphics[width=0.95\linewidth]{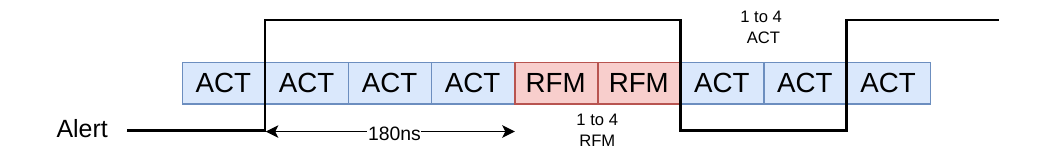} 
    \caption{ABO alerts sent by the DRAM must be separated by ACT commands}
    \label{fig:abo}
\end{figure}

The PRAC-N counters of rows refreshed through an \ac{RFM} command are themselves incremented, potentially triggering additional mitigative refreshes. Each \ac{RFM} command stalls the DRAM for a duration of $\mathit{t_{RFM}}$. To prevent memory accesses from being completely blocked by consecutive mitigations, the DRAM must receive at least $N$ (1, 2, or 4) \ac{ACT} commands before issuing another \ac{ABO} alert \cite{jedec_prac}. Figure~\ref{fig:abo} illustrates the behaviour of such an \ac{ABO} alert.

Hardware countermeasures are effective when the RowHammer detection threshold ($\mathit{T_{RH}}$) (i.e., the number of activations after which the first RowHammer-induced bit flip would occur) remains sufficiently high (e.g., 1000 row activations before trigger~\cite{break}). However, as $\mathit{T_{RH}}$ decreases with technology scaling, mitigations must intervene more frequently, increasing memory interference and degrading overall system performance. This effect has led to the emergence of performance-oriented attacks~\cite{rogue_rfm, not_so_ref, moat}, which deliberately trigger repeated mitigative actions to stall the memory subsystem and disrupt normal operation. To date, the impact of such attacks has primarily been evaluated in terms of average performance degradation in general-purpose systems. However, their consequences on the worst-case execution time of real-time workloads have not yet been thoroughly investigated. Investigating this impact is the main motivation behind our work, leading us to formulate the research question found at the end of the next subsection.

\subsection{DRAM in Critical Real-Time Systems}

In safety-critical real-time systems, predictability is prioritised over average-case performance. Consequently, several optimisations commonly used in general-purpose systems are either restricted or replaced. For instance, open-row DRAM policies exploit spatial locality to improve throughput, but they also introduce highly data-dependent access latencies that complicate \ac{WCET} analysis. Real-time \acp{MC} therefore frequently rely on closed-row policies, which provide more deterministic timing behaviour at the cost of reduced average-case performance~\cite{close_row}. 

In this context, ensuring predictable worst-case memory access latencies is essential. However, RowHammer and its countermeasures have received very little attention in the context of real-time systems. A recent work~\cite{Guha_Chakrabarti_2022} studied RowHammer attacks in shared cloud FPGA environments hosting real-time tasks. The proposed approach focuses on protecting task execution against corruption through \textit{reactive} fault-tolerance techniques, namely spatial redundancy with majority voting and task re-execution after error detection. As such, the objective is to tolerate or recover from faults once bit flips have already occurred. In contrast, modern DRAM systems increasingly rely on \textit{preventive} RowHammer countermeasures implemented directly within memory devices or memory controllers~\cite{survey_rh_mitig}, such as PRAC in recent DDR5 memories, which proactively trigger mitigative actions \textit{before faults occur}. These mechanisms fundamentally differ from reactive redundancy schemes because they directly interfere with the timing behaviour of memory accesses during normal execution. Since such countermeasures are already deployed in recent commercial DRAM devices (e.g., PRAC), understanding their impact on the timing behaviour of real-time systems is therefore paramount.

In the described context, this work formulates and answers the following research question:
\vspace{-0.5em}
\begin{researchquestion}
    Can RowHammer counter-based countermeasures impact, and to what extent, the WCET of typical safety-critical real-time systems workloads?
\end{researchquestion}

\section{System Model}
\label{sec:syst_model}

\begin{figure}
    \centering
    \includegraphics[trim=20pt 10pt 20pt 10pt, clip,width=0.75\linewidth]{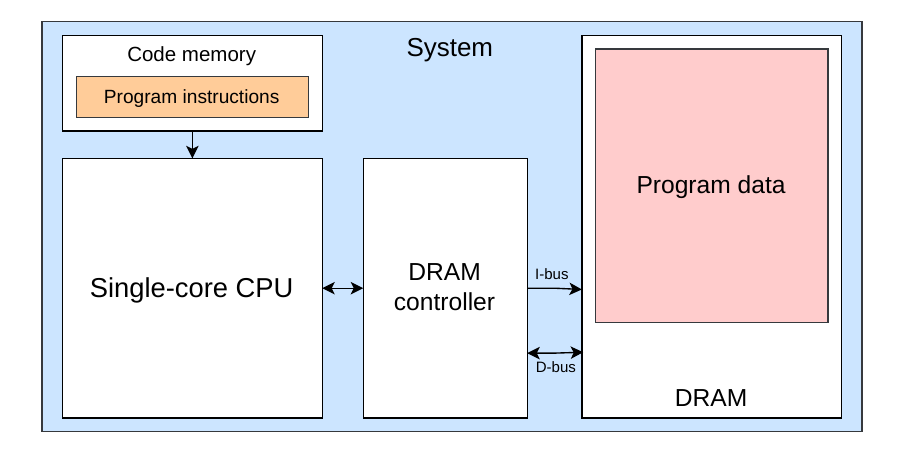}
    \caption{Architecture of the system used in this work}
    \label{fig:sys1}
    \vspace{-1em}
\end{figure}

We consider a single-core real-time system running bare-metal code as depicted in \figurename~\ref{fig:sys1} to illustrate the problem. The processor features a Harvard architecture, with separate instruction and data memories. The system relies on a DRAM memory for data storage, such as DDR5. The DRAM is protected against RowHammer using counter-based countermeasures, such as PRAC-N countermeasure (Section~\ref{sec:rh_mitig}), while the \ac{MC} follows a closed-row policy\footnote{The analysis and observations presented in this work can be extended to memory controllers implementing an open-row policy. In such systems, the number of row activations (\acp{ACT}) is expected to be lower due to row-buffer locality, which would likely reduce the impact of RowHammer mitigation mechanisms. Nevertheless, the same fundamental interactions between the memory access pattern and the mitigation strategy still apply.}.

Our objective is to evaluate the impact of PRAC-N on the execution time of genuine workloads in this configuration by analysing program executions under different initial memory states. Such initial states may result from software executed prior to the considered workload, including potentially malicious code. This assumption is increasingly relevant as modern safety-critical embedded systems are no longer fully isolated and are progressively exposed to external connectivity and associated attack vectors \cite{bloomfield_2013_SP_secu_safety}.

This system model captures several design choices that are representative of safety-critical real-time systems, where predictability is often prioritised over average-case performance. In particular, cacheless designs \cite{no_cache} and closed-row memory policies \cite{close_row} are commonly used to simplify timing analysis and enable tighter \ac{WCET} bounds. Moreover, a single-core system can be considered the lower-bound scenario for memory access contention. Demonstrating a negative impact on execution time for such a system would be equally applicable to systems with a larger number of cores and thus higher contention, which could make this negative impact even worse.

\section{The \Jenga Attack}
\label{sec:jenga}

\subsection{System Setup and Threat Model}
\label{sec:jenga_setup}

The task model~\cite{Sib14} is a set of sporadic tasks, \{$\tau$\}, where each task $\tau_i \in \tau$ has the parameters: \{$P_i$, $C_i$, $D_i$\},
where $P_i$ is the period, $C_i^0$ is the worst-case execution time and
$D_i$ is the deadline, with $D_i \leq P_i$. We assume a preemptive fixed-priority schedule with a set of real-time priorities, \{$\mathit{Pri}$\}
are fixed such that, $\forall \tau_i, \tau_j \in \{\tau\}$ and $pri_{\tau_i}$, $pri_{\tau_j} \in \{Pri\}$, then either $pri_{\tau_i} \prec pri_{\tau_j}$, if $\tau_i$ has a higher priority than $\tau_j$ or $pri_{\tau_i} \succ pri_{\tau_j}$
if $\tau_j$ has a higher priority than $\tau_i$.
Of course, it is also possible that $pri_{\tau_i}$ and $pri_{\tau_j}$ share the same priority level. For illustration reasons, we consider two tasks, $\tau_{1}$, which is the victim, and $\tau_{2}$, which is the attacker with the lowest priority. 
A summary of the notations used throughout this section is provided in Table~\ref{tab:notation}.

\begin{table}[H]
    \centering
    \setlength{\tabcolsep}{2pt}
    \begin{tabular}{l|p{0.8\linewidth}}
         $\tau_i$ & Task $i$ \\
         $C_i^0$ & Refresh-unaware worst-case execution time of $\tau_i$ \\
         $C_i^{REF}$ & Contribution of periodic refreshes to the WCET of $\tau_i$ \\
         $C_i^{mitig.}$ & Contribution of mitigative actions to the WCET of $\tau_i$ \\
         $n_i$ & Number of memory accesses performed by $\tau_i$ \\
         $WCET_i$ & Refresh- and countermeasure-aware WCET of $\tau_i$ \\
         $n_{RFM}$ & Number of aggressor rows whose neighbouring rows are refreshed by an \ac{RFM} command \\
         $T_{RH}$ & RowHammer detection threshold \\
         $r_{bank}$ & Number of rows in a DRAM bank \\
         $N$ & Number of \ac{RFM} commands per ABO alert in PRAC-N \\
         $n_{ACT}^{row_j}$ & Number of ACT commands targeting row $j$ \\
         $t_{REFI}$ & Time interval between two consecutive REF commands \\
         $t_{REF}^{max}$ & Worst-case latency between the reception of a REF command by the DRAM and the completion of the refresh operation \\
         $t_{RFM}^{\CR{max}}$ & Worst-case latency between the reception of an RFM command by the DRAM and the completion of the refresh operation \\
         \CR{$t_{REFW}$} &\CR{Time window over which all DRAM rows are refreshed once}
    \end{tabular}
    \caption{List of notations}
    \label{tab:notation}
\end{table}

We consider an attacker whose objective is to delay $\tau_{1}$ as long as possible, potentially causing it to miss its deadline in a given execution instance. The attacker controls task $\tau_{2}$ and can perform memory accesses. We assume that $\tau_{1}$ and $\tau_{2}$ do not share data and do not communicate directly. The attacker has no additional privileges beyond controlling the execution behaviour and memory access pattern of $\tau_{2}$. The attack, therefore, relies solely on indirect interference via the internal state of the DRAM RowHammer mitigation mechanism. We additionally assume that the attacker knows the state of the memory, i.e., the value of the PRAC-N counters, as well as the conditions under which an \ac{ABO} alert is triggered, including the activation threshold $T_{RH}$. Previous work has already demonstrated how refresh operations due to mitigative actions can be observed from the user space to build side and covert channels~\cite{bostanci_UnderstandingMitigating_CCSCA_2025}.

\subsection{The Attack}
\label{sec:jenga_attack}

\subsubsection{Overall Idea}
\label{sec:cascade}

\begin{figure}[t]
    \centering

    \begin{subfigure}[t]{0.115\textwidth}
        \centering
        \includegraphics[width=\linewidth]{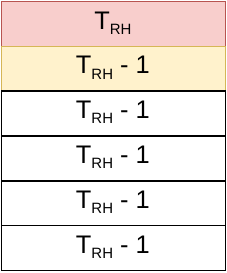}
        \caption{\footnotesize A row activation exceeds the PRAC threshold, triggering an alert and an \ac{RFM} refresh of its neighbour.}
    \end{subfigure}
    \hfill
    \begin{subfigure}[t]{0.115\textwidth}
        \centering
        \includegraphics[width=\linewidth]{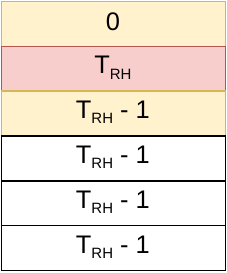}
        \caption{\footnotesize The refreshed row is activated, its counter exceeds $T_{RH}$, and a new alert is triggered.}
    \end{subfigure}
    \hfill
    \begin{subfigure}[t]{0.115\textwidth}
        \centering
        \includegraphics[width=\linewidth]{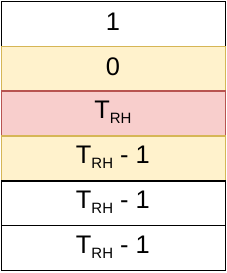}
        \caption{\footnotesize The alert propagates to the next row, which is refreshed and activated in turn.}
    \end{subfigure}
    \hfill
    \begin{subfigure}[t]{0.115\textwidth}
        \centering
        \includegraphics[width=\linewidth]{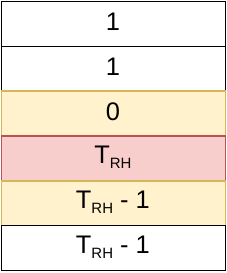}
        \caption{\footnotesize Alerts propagate until all rows have been refreshed by an \ac{RFM} command.}
    \end{subfigure}

    \caption{Example of an \ac{RFM} cascade over time. In each subfigure, the red row acts as the aggressor row that triggers the \ac{RFM} mechanism. Consequently, its neighbouring rows (in yellow) are refreshed.}
    \label{fig:cascade}
\end{figure}

As mentioned in Section \ref{sec:rh_mitig}, PRAC-N can exhibit a cascading behaviour when multiple neighbouring rows have activation counters close to the RowHammer detection threshold $T_{RH}$~\cite{pvac}. In such a situation, when the PRAC mechanism triggers an \ac{RFM} mitigative command on one aggressor row, the neighbouring rows, whose counters may be close to saturation, are refreshed. Since these neighbouring, refreshed rows are considered activated by the PRAC mechanism \cite{pvac}, their counters increment, potentially exceeding $T_{RH}$ and therefore becoming eligible to trigger additional alerts. Consequently, a single alert can propagate through multiple rows and generate a cascade of mitigative refreshes, as illustrated in \figurename~\ref{fig:cascade}. However, this \textit{tower} of pending mitigations does not collapse on its own. Due to the behaviour of the ABO mechanism illustrated in \figurename~\ref{fig:abo}, the memory controller must issue $N$ ACT commands after an alert before a new alert can be generated. As a result, the \textit{fall of the tower} depends on \textit{normal} memory accesses: the DRAM is not completely stalled while all rows are refreshed at once, but instead experiences short bursts of mitigation-induced stalls interleaved with regular memory accesses, roughly every $N$ ACT commands.

The \Jenga attack\footnote{The name comes from the Jenga game. The attacker progressively builds the ``tower'' by manipulating the PRAC counter state, and eventually performs the access that causes the cascade of mitigations.} exploits this cascading behaviour. When it executes, the attacker-controlled task $\tau_2$ repeatedly accesses neighbouring memory rows in its memory space in order to progressively increase their PRAC counters. More precisely, the attacker performs enough activations to bring selected counters just below the activation threshold $T_{RH}$. The objective is to prepare a \Jenga tower sufficiently high such that its collapse maximises the execution time of $\tau_{1}$ when it subsequently executes. In a final access, the attacker deliberately triggers the first alert, initiating the cascade of mitigations and leaving the DRAM in a highly adverse state for the next task to execute.

\subsubsection{Manipulating the Memory State Across Multiple Executions}

Because $\tau_2$ executes only in the background and may be preempted by $\tau_1$, the attacker may not be able to construct a sufficiently large \Jenga tower during a single uninterrupted execution interval. Nevertheless, since the PRAC counter state persists across preemptions and resumptions of $\tau_2$, the attacker can progressively prepare the memory state over multiple execution intervals. By repeatedly extending the \Jenga tower each time $\tau_2$ resumes execution, the attacker can eventually create a memory configuration that induces significant perturbations in the execution of $\tau_1$.

\subsubsection{Effect of Periodic REF Commands}

The memory controller periodically issues REF commands to the DRAM. These commands refresh groups of rows and therefore increment their corresponding PRAC counters. This behaviour complicates the attacker's manipulation of the memory state, as periodic refreshes may modify counters independently of the attacker's actions and may trigger the fall of the \Jenga tower prematurely. In the current work, we do not explicitly model the impact of REF commands and assume that the attacker can compensate for their effects by appropriately adjusting the number of accesses performed to each row. We further discuss the implications of this assumption in Section~\ref{sec:discussion}.

\subsection{Expected Impact of \Jenga}

The additional execution time incurred by $\tau_1$ can be estimated by bounding the number of extra mitigations triggered when the \Jenga tower collapses. Let us assume that the attacker has successfully prepared $k$ rows whose PRAC counters are equal to $T_{RH}-1$ before collapsing the tower prior to $\tau_1$ execution. Consequently, the number of additional \ac{RFM} commands issued by the \ac{MC} is bounded by:
$$
    \ceil{\frac{k}{n_{RFM}}}
$$
where $n_{RFM}$ denotes the number of aggressor rows mitigated by a single \ac{RFM} command.

Additionally, alerts cannot be triggered back-to-back indefinitely. As discussed previously and illustrated in \figurename~\ref{fig:abo}, genuine memory accesses must be performed between two successive alerts. More precisely, whether $N=1$, $2$, or $4$, the memory controller must issue at least $N$ \ac{ACT} commands before another burst of $N$ \ac{RFM} commands can be triggered. Consequently, the number of additional \ac{RFM} commands generated during the execution of $\tau_1$ is also bounded by:
\begin{equation}
    n_{ACT}
\end{equation}
where $n_{ACT}$ denotes the number of \ac{ACT} commands issued during the execution of $\tau_1$. By construction, we have $n_{ACT} \leq n_1$, with $n_{ACT} = n_1$ under a strict closed-row policy (i.e. cap = 1).

Finally, the additional execution time induced by the attack can be bounded by:
\begin{equation}
    \min\left(
    \ceil{\frac{k}{n_{RFM}}},
    n_{ACT}
    \right)
    \cdot t_{RFM}^{\CR{max}}
\end{equation}
where $t_{RFM}^{\CR{max}}$ denotes the worst-case latency induced by a single \ac{RFM} command.

Note that this bound is expected to overestimate the impact of the attack in some practical scenarios. Indeed, we conservatively assume that no \ac{ACT} command is issued during the 180ns interval separating the reception of an \ac{ABO} alert from the transmission of the first \ac{RFM} command.

\section{WCET computation}
\label{sec:wcet}

Since mitigative actions can invalidate WCET estimates that do not account for RowHammer countermeasures, we provide a theoretical analysis of the impact of mitigative operations on the execution time of programs running in our system model.

\subsection{Unprotected DRAM WCET}

As they account for a negligible part of the memory latency, periodic DRAM refreshes are often ignored when determining the WCET of a program \cite{RTCSA_2024, ref_ignored_1, ref_ignored_2, ref_ignored_3, ref_ignored_4}. However, a few works aim at characterising the impact of these refreshes \cite{first_ref, predictable_ref, WCET_ref}. Notably, the periodic refresh-aware WCET of a program executing \textbf{without interruption} can be written as:
\begin{equation} \label{eq:Cref}
    WCET = C^0 + C^{REF} = C^0 + \ceil{\frac{C^0}{t_{REFI} - t_{REF}^{max}}} \cdot t_{REF}^{max}
\end{equation}
where $C^0$ is the WCET of the program without taking periodic refreshes into account, $t_{REFI}$ is the time interval between two REF commands and $t_{REF}^{max}$ is the maximum delay a refresh operation can take. Previous work \cite{WCET_ref} has formulated an equation to derive $t_{REF}^{max}$ from the DRAM timing parameters.

\subsection{Protected DRAM WCET}
\label{sec:wcet_analysis}

Previous works studying the impact of DRAM refreshes on WCET only consider periodic refreshes. In this case, determining the number of REF commands issued during a time interval is relatively straightforward. However, mitigative actions are event-driven rather than periodic, making their analysis significantly more challenging.

Similarly to previous work \cite{first_ref, predictable_ref, WCET_ref}, we decompose the WCET of a program as follows:
\begin{equation}
\label{eq:decomp_wcet}
    WCET = C^0 + C^{mitig.} + C^{REF}
\end{equation}
where $C^0$ is assumed to be known.

\begin{figure}
    \centering
    \includegraphics[trim=0pt 4pt 6pt 0pt, clip,width=0.82\linewidth]{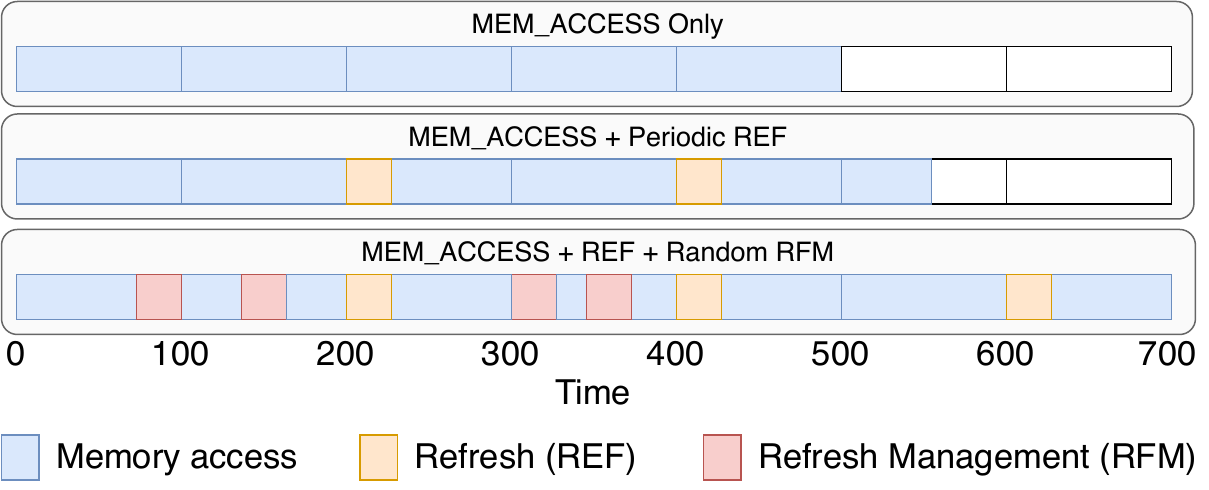}
    \caption{Qualitative impact of \ac{RFM} commands on the number of REF commands issued to the DRAM.}
    \label{fig:REF_and_RFM}
\end{figure}

In this context, equation (\ref{eq:Cref}) no longer holds directly. As illustrated in Figure~\ref{fig:REF_and_RFM}, mitigative actions (\ac{RFM}) increase execution time, which in turn may trigger additional periodic refreshes (REF) during execution. \CR{Additionally, periodic refreshes only contribute to execution time when they overlap with memory requests and stall their execution. Therefore, the number of refresh-induced stalls is bounded by the maximum number of memory requests $n_{mem}$ performed by the program.} Consequently, the refresh-related contribution becomes:
\begin{equation}
\label{eq:cref_with_mitig}
    \CR{C^{REF} = \min\left(\ceil{\frac{C^0 + C^{mitig.}}{t_{REFI} - t_{REF}^{max}}}, n_{mem}\right) \cdot t_{REF}^{max}}
\end{equation}

The remaining challenge is therefore to compute $C^{mitig.}$. While the exact derivation depends on the mitigation mechanism under consideration, we propose a novel  methodology for deterministic counter-based countermeasures:

\begin{enumerate}
    \item Determine the worst-case initial memory state, i.e. the value of each mitigation counter that maximises the number of mitigative actions. To obtain a tighter bound, only states reachable under the considered system and attacker models can be considered.
    
    \item Simulate the execution of the task and count the number of mitigative actions triggered at runtime. 
    
    \item Determine the worst-case latency induced by a mitigative action.\footnote{Some mechanisms also modify the latency of accesses that do not trigger mitigative actions (e.g., to support the update of the counters, $t_{RC}$ in PRAC-defended DRAMs increased from 46ns to 52ns \cite{salt}). Such effects should be included in $C_i^0$, not in $C^{mitig.}$.}
    
    \item Multiply the maximum number of mitigative actions by the worst-case latency of a single action.
\end{enumerate}

We now apply this methodology to the PRAC-N countermeasure, which is \CR{standardised for} recent DDR5 DRAM devices. We assume that the genuine program executes without interruption. We further assume that the only mechanism capable of decreasing the PRAC counter of a row is the refresh of neighbouring rows through an \ac{RFM} command. In other words, REF commands do not reduce activation counters, implying that $C_i^{mitig.}$ is independent from $C_i^{REF}$. While this assumption may depend on vendor-specific PRAC implementations and may therefore not hold for all DRAM modules, previous work supports this hypothesis \cite{pvac}.
We denote by $n_{RFM}$ the number of aggressor rows whose neighbouring rows are refreshed by a single \ac{RFM} command, i.e., the number of counters reset by an \ac{RFM} command.
We first determine the worst-case initial memory state before the program's execution. For PRAC-N, this corresponds to all counters being initialised to $T_{RH}-1$. Such a state enables cascades of mitigations as presented in section \ref{sec:cascade}.
In this scenario, the total number of alerts raised during task execution can be bounded by the sum of:
\begin{itemize}
    \item $PRAC_{init}$, the number of alerts caused by the initial memory state;
    \item $PRAC_{pattern}$, the number of alerts genuinely triggered by the task memory access pattern, assuming all counters initially start from zero.
\end{itemize}

We first derive a bound for $PRAC_{init}$. Two independent phenomena limit the number of cascaded alerts.

First, a memory bank contains a finite number of rows, denoted $r_{bank}$. Since each triggered alert leads to issuing $N$ \ac{RFM} commands, and each \ac{RFM} command resets $n_{RFM}$ counters, we obtain:
\begin{equation}
\label{eq:majorant_1}
    PRAC_{init} \leq \ceil{\frac{r_{bank}}{N \cdot n_{RFM}}}
\end{equation}
Second, as illustrated in \figurename~\ref{fig:abo}, the memory controller cannot issue more than $N$ consecutive \ac{RFM} commands. It must send at least $N$ ACT commands between two alerts. Although some ACTs may be processed while the MC receives the ABO alert and prepares RFMs, it is also possible that it issues no commands during this interval. We therefore conservatively assume that each burst of \ac{RFM} commands must be separated by at least $N$ ACT commands. This yields:
\begin{equation}
\label{eq:majorant_2}
     PRAC_{init} \leq \ceil{\frac{n_{ACT}}{N}}
\end{equation}
Combining equations \ref{eq:majorant_1} and \ref{eq:majorant_2} gives:
\begin{equation}
\label{eq:PRAC_init}
    PRAC_{init} \leq min\left( \ceil{\frac{r_{bank}}{N \cdot n_{RFM}}}, \ceil{\frac{n_{ACT}}{N}} \right)
\end{equation}
\CR{To account for limitations on the size of the \Jenga tower that an attacker can construct (e.g., due to memory partitioning constraints), we can replace the $r_{bank}$ term by the \textit{attacker's budget} $r_{attacker}$, which denotes the number of rows that the attacker can prepare. We have $r_{attacker} \leq r_{bank}$.}\\

We now derive a bound for $PRAC_{pattern}$. This quantity depends on the task memory access pattern and can be bounded by the number of times each row is activated more than the PRAC detection threshold $T_{RH}$:
\begin{equation}
\label{eq:PRAC_pattern}
    PRAC_{pattern} \leq \sum_{j \in rows} \floor{\frac{n_{ACT}^{row_j}}{T_{RH}}}
\end{equation}

Combining equations \ref{eq:PRAC_init} and \ref{eq:PRAC_pattern}, the total number of PRAC alerts is bounded by:
\begin{equation}
\label{eq:prac}
     n_{PRAC} \leq min\left( \ceil{\frac{r_{bank}}{N \cdot n_{RFM}}}, \ceil{\frac{n_{ACT}}{N}} \right) + \sum_{j \in rows} \floor{\frac{n_{ACT}^{row_j}}{T_{RH}}}
\end{equation}

Since each alert causes the memory controller to issue $N$ \ac{RFM} commands, the mitigation-related WCET contribution becomes:
\begin{equation}
\label{eq:boound_cprac}
     C_i^{PRAC} \leq N \cdot n_{PRAC}^{max} \cdot t_{RFM}^{\CR{max}}
\end{equation}
where $n_{PRAC}^{max}$ is given by the right-hand side of equation \ref{eq:prac}.

Finally, equations \ref{eq:decomp_wcet}, \ref{eq:cref_with_mitig}, and \ref{eq:boound_cprac} together provide a WCET bound for tasks executing on DRAM protected by PRAC-N.

\section{Experimental Evaluation}
\label{sec:expe}

We conduct two complementary sets of experiments. 
In the first experimental set, we implement the attack scenario described in Section \ref{sec:jenga_setup} and \ref{sec:jenga_attack} for three different workloads that from the TACLeBench benchmark suite  (see Section~\ref{sec:TACLeBench}), which are executed periodically. We evaluate the impact of \Jenga on the execution time of genuine tasks as we vary the number of rows whose PRAC counters can be manipulated by the attacker. The results for this experiment are shown in Section~\ref{sec:results:superloop}.
In the second experimental set, we study the influence of the initial PRAC counter state on program execution time. To this end, we execute several genuine TACLeBench workloads under two different initial memory states: (i) a clean state where all PRAC counters are initialised to zero, and (ii) an \textit{adversarial} state where all counters are initialised to $T_{RH}-1$, quantifying the impact of a fully successful attack for each benchmark. These results are shown in Section~\ref{sec:results:initmemstate}.
All experiments are conducted using the Ramulator 2.0 DRAM simulator integrated with gem5 in order to model a complete system stack. \CR{The workload binaries are compiled using the RISC-V GNU Compiler and version 15.2.0 of GCC}. The simulation parameters are summarised in Table \ref{tab:param_simul}.

\subsection{The gem5 Simulator}

gem5 \cite{gem5, gem5bis} is a modular, event-driven, cycle-level simulator widely used for computer architecture research. It provides multiple configurable components (CPU models, memory systems, interconnects) that can be composed to simulate a full system.

In our experiments, we use a \texttt{RiscvTimingSimpleCPU}, which models an in-order RISC-V core with timing annotations. The processor is connected to two distinct memory components: (i) an instruction memory, modelled as a fast local memory (analogous to a tightly-coupled SRAM), and (ii) a data DRAM memory backed by Ramulator 2.0. 

We run gem5 in full-system (FS) mode. However, instead of booting a full operating system, we execute bare-metal code within the simulated environment. This approach allows us to instantiate a system without OS-level abstractions, in particular, avoiding virtual memory translation. As a result, we have full control over DRAM accesses.

\subsection{Ramulator 2.0 simulator}

Ramulator 2.0 \cite{ramulator2} is a cycle-accurate DRAM simulator designed to model modern memory standards and is widely used in RowHammer studies, particularly for evaluating the effectiveness and impact of countermeasures on system performance.
In our setup, gem5 is interfaced with Ramulator such that all data memory requests generated by the CPU are forwarded to the DRAM model. This enables detailed modelling of DRAM timing constraints and command behaviour.
We modified parts of Ramulator 2.0 to better capture some DRAM-level effects that are abstracted in the default implementation. In particular, we found that \ac{RFM} (Refresh Management) commands did not update the PRAC counters associated with refreshed victim rows, which is inconsistent with the behaviour described in \cite{pvac}. Without this mechanism, the effects of the \Jenga attack cannot be correctly modelled, as the interaction between refresh operations and activation counters is essential to its effectiveness. We will release these modifications to Ramulator along with \Jenga code.

\subsection{TACLeBench Workloads}
\label{sec:TACLeBench}

To evaluate the impact of the initial DRAM state on program execution time, we used TACLeBench benchmark suite's workloads~\cite{tacle}. TACLeBench is widely used in the real-time systems community for evaluating \ac{WCET} analysis techniques, as it provides self-contained and processor-independent embedded workloads. The suite gathers and categorises benchmarks into kernel, sequential, application, test, and parallel benchmarks. In our study, we focused on the kernel and sequential benchmark categories, as they provide representative embedded workloads with diverse execution characteristics and memory access patterns while remaining suitable for timing analysis experiments. In particular, the selected benchmarks exhibit varying memory footprints and memory intensities, ranging from computation-intensive workloads to more memory-intensive applications. The benchmarks were executed under different initial memory states in order to quantify the impact of mitigative refresh mechanisms on execution time.


\CR{Additionally, we apply the PRAC-aware WCET analysis presented in Section~\ref{sec:wcet_analysis} to several TACLeBench workloads. Since the baseline execution time $C^0$ is a theoretical quantity that cannot be easily observed on a real system (e.g. disabling refreshes may not be possible), we instead estimate the probabilistic WCET without the attacker, while keeping the RowHammer countermeasure enabled. We denote this baseline as $pWCET^0$ and estimate it using a Measurement-Based Probabilistic Timing Analysis (MBPTA) approach. More precisely, we rely on the block-maxima method~\cite{measurement_based_wcet, romaric} and fit a Gumbel distribution to the observed execution times.}

\CR{For each workload, we perform 650 consecutive executions without resetting the DRAM state between runs while keeping PRAC enabled. Since the workloads operate on identical input data across all runs, execution-time variations are not caused by data-dependent behaviour. Instead, the observed variability originates solely from DRAM-related effects, namely the timing of periodic refresh operations and the evolution of the PRAC counter state between consecutive executions.}

\CR{The attacker-aware WCET bound is then obtained by instantiating the theoretical model with this measured baseline:
\begin{equation}
    pWCET^{0} = C^{0} + C^{0}_{PRAC} + C^{0}_{REF}
\end{equation}
with 
\[
\left\{
    \begin{array}{l}
         C^{0}_{PRAC} = PRAC_{pattern} \cdot N \cdot t_{RFM}^{max}\\
         C^{0}_{REF} = \lceil \frac{C^{0} + C^{0}_{PRAC}}{t_{REFI} - t_{REF}^{max}}\rceil \cdot t_{REF}^{max} 
    \end{array}  
\right.
\]
and 
\begin{equation}
    WCET = pWCET^{0} + C^{attack}_{PRAC} + C^{attack}_{REF}
\end{equation}
where 
\[
\left\{
    \begin{array}{l}
        C^{attack}_{PRAC} = PRAC_{init} \cdot N \cdot t_{RFM}^{max} \\
        C^{attack}_{REF} = \lceil \frac{C^{attack}_{PRAC}}{t_{REFI} - t_{REF}^{max}} \rceil \cdot t_{REF}^{max}
    \end{array}
\right.
\]
}

\begin{table}[]
\caption{Simulation parameters}
\resizebox{\columnwidth}{!}{%
\begin{tabular}{l|l}
\hline
\textbf{Processor} & \begin{tabular}[c]{@{}l@{}}Single Core RiscvTimingSimpleCPU, \\ in-order, 1GHz clock frequency\end{tabular} \\ \hline
\textbf{Code + Stack Memory} & 4ns latency, SRAM-like memory \\ 
\hline
\textbf{Data Memory} & \begin{tabular}[c]{@{}l@{}}DDR5\_3200AN, 8GiB, 1 channel, \\ 1 rank, x4, 8 bank groups, \\ 2 banks/bank group, 65536 rows/bank\end{tabular} \\ \hline
\textbf{Data Memory Controller} & \begin{tabular}[c]{@{}l@{}}Closed-row policy (cap = 4),\\ FCFS scheduling, all-bank \\ refresh policy, $n_{RFM} = 1$, \\ PRAC-N with $N = 1$ and $T_{RH} = 128$\end{tabular} \\ \hline
\end{tabular}%
}
\label{tab:param_simul}
\vspace{-1em}
\end{table}

\section{Results}
\label{sec:results}

\begin{figure*}[h!]
    \centering

    \begin{subfigure}[t]{0.32\textwidth}
        \centering
        \includegraphics[width=\linewidth, trim={17pt 20pt 17pt 15pt}, clip]{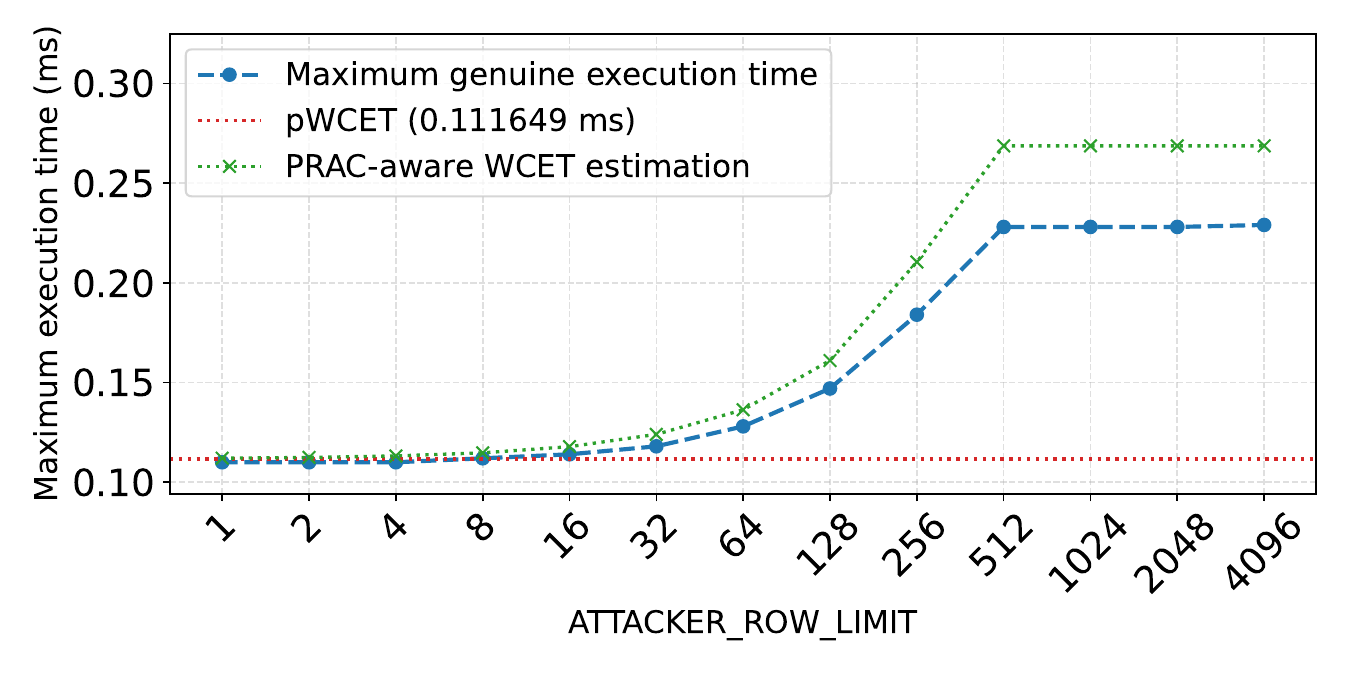}
        \caption{\footnotesize CountNegative workload}
    \end{subfigure}
    \hfill
    \begin{subfigure}[t]{0.32\textwidth}
        \centering
        \includegraphics[width=\linewidth, trim={17pt 20pt 17pt 17pt}, clip]{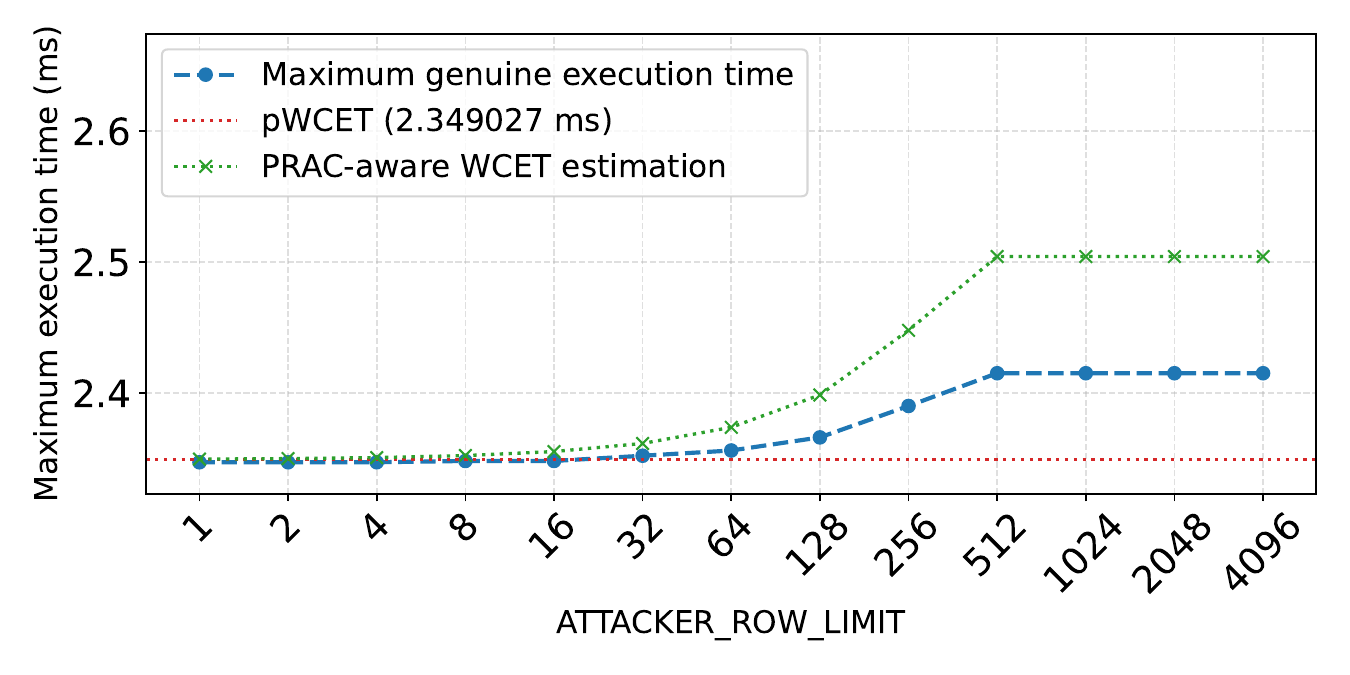}
        \caption{\footnotesize LMS workload}
    \end{subfigure}
    \hfill
    \begin{subfigure}[t]{0.32\textwidth}
        \centering
        \includegraphics[width=\linewidth, trim={17pt 20pt 17pt 15pt}, clip]{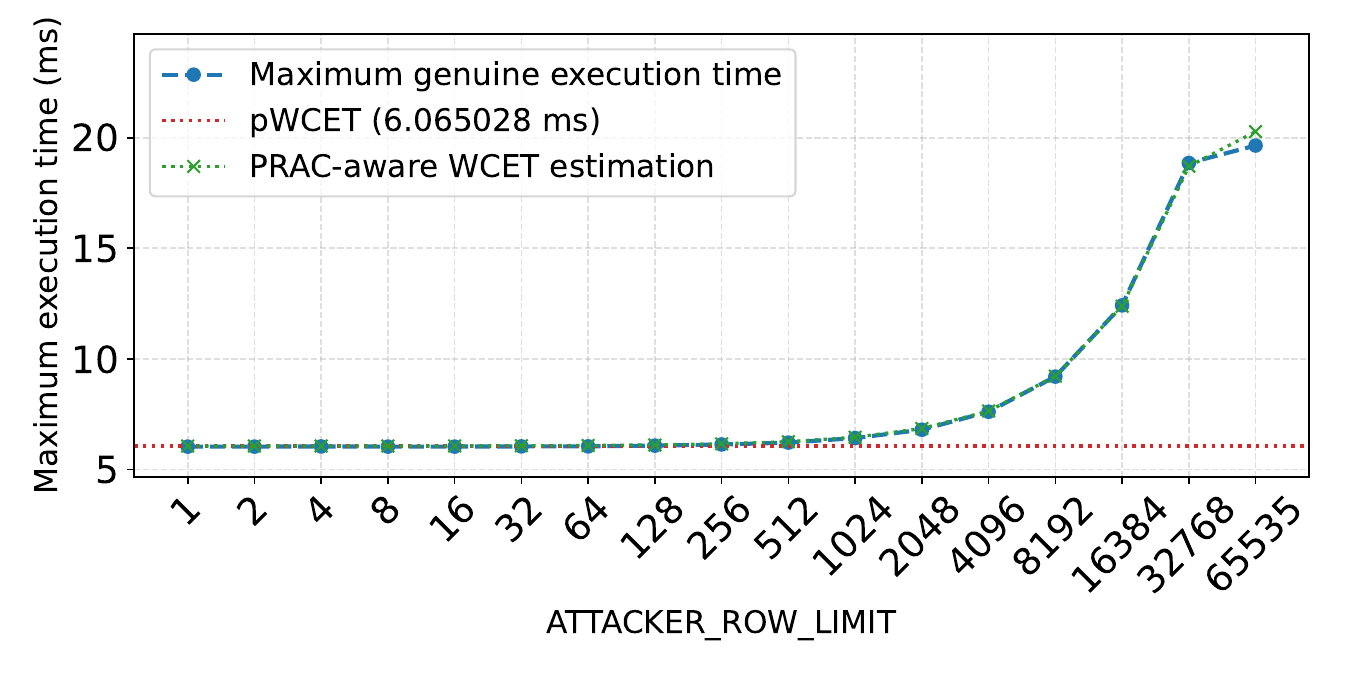}
        \caption{\footnotesize FFT workload}
    \end{subfigure}

    \begin{subfigure}[t]{0.32\textwidth}
        \centering
        \includegraphics[width=\linewidth, trim={17pt 20pt 17pt 15pt}, clip]{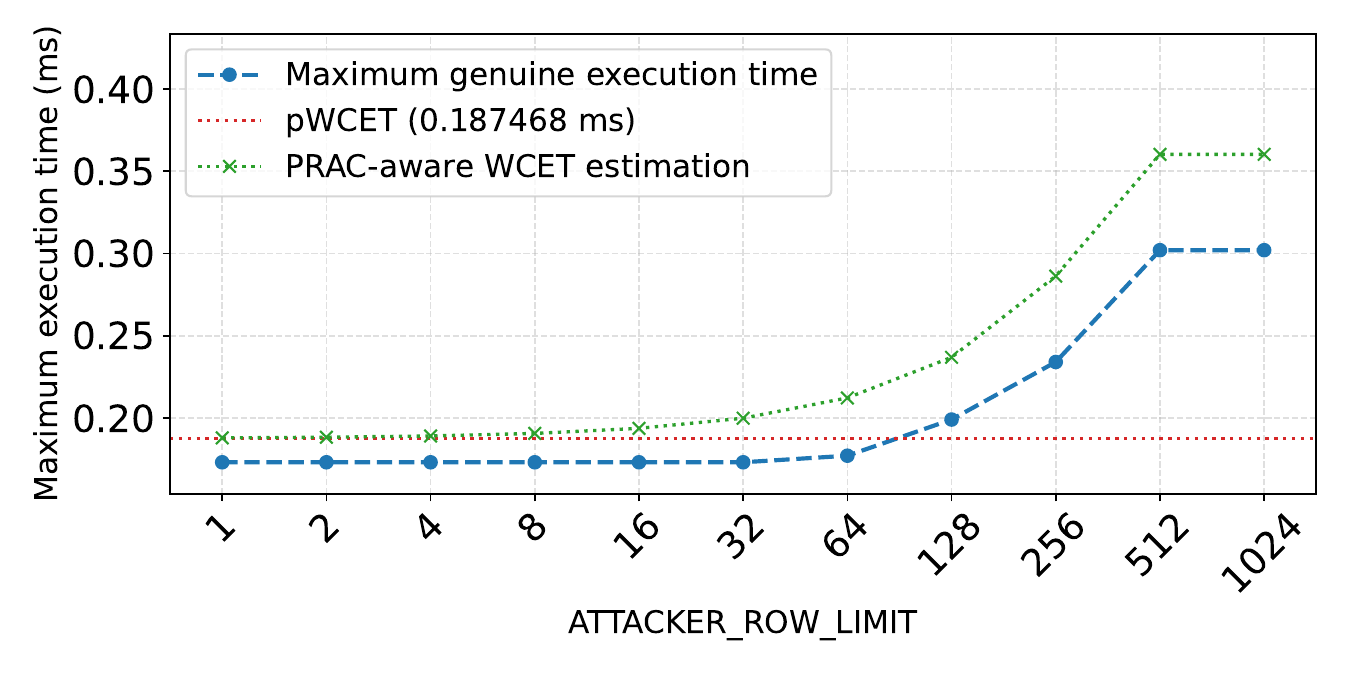}
        \caption{\footnotesize Bitcount workload}
    \end{subfigure}
    \hfill
    \begin{subfigure}[t]{0.32\textwidth}
        \centering
        \includegraphics[width=\linewidth, trim={17pt 20pt 17pt 15pt}, clip]{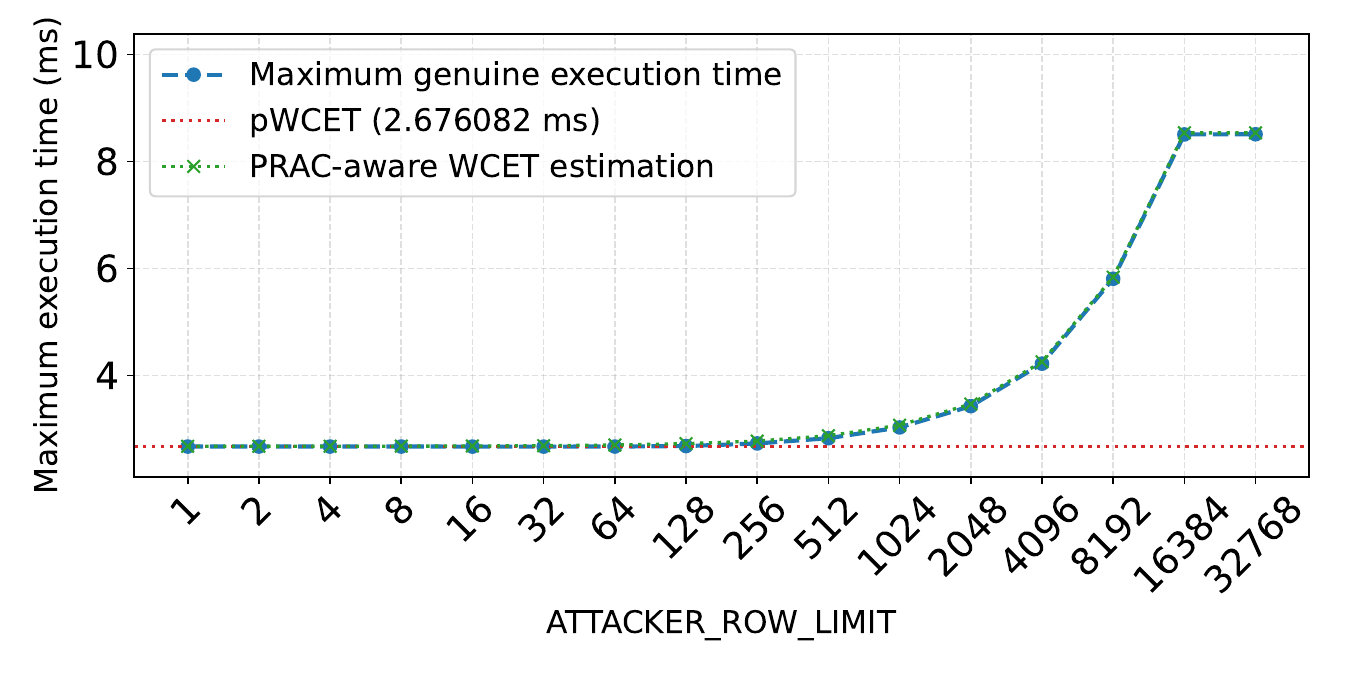}
        \caption{\footnotesize Bitsort workload}
    \end{subfigure}
    \hfill
    \begin{subfigure}[t]{0.32\textwidth}
        \centering
        \includegraphics[width=\linewidth, trim={17pt 20pt 17pt 15pt}, clip]{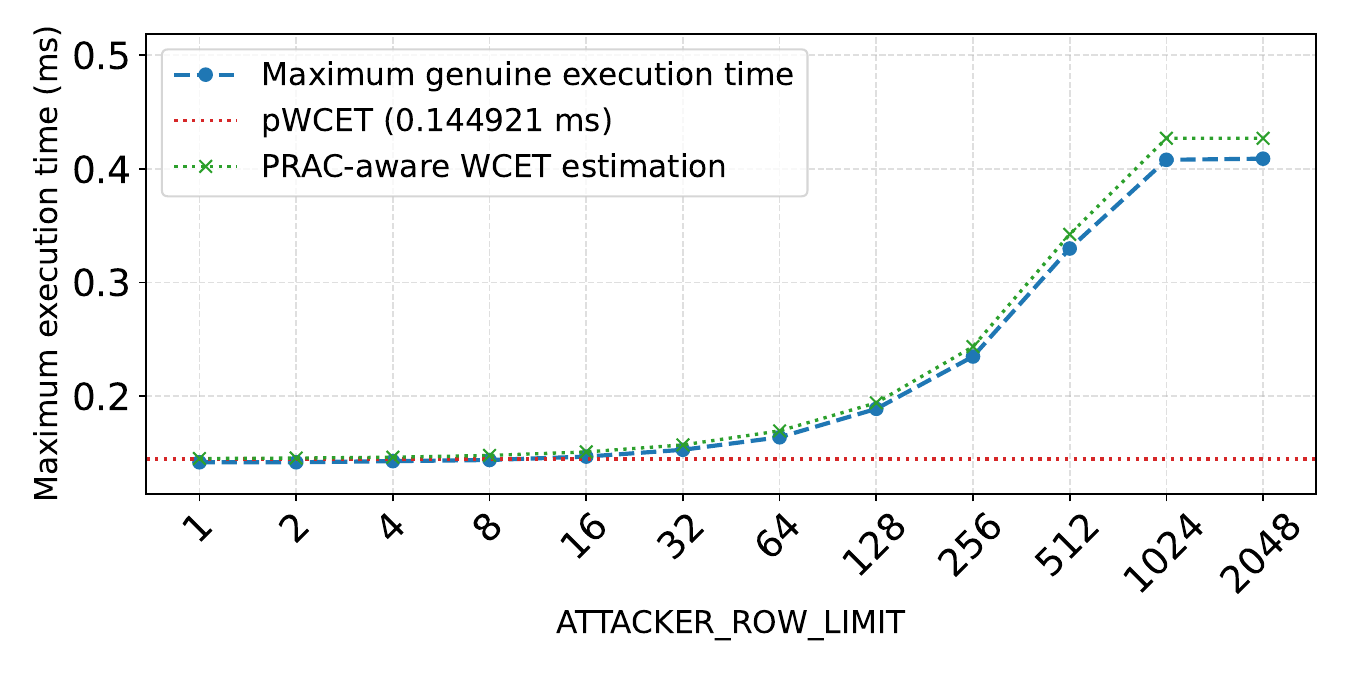}
        \caption{\footnotesize Deg2Rad workload}
    \end{subfigure}

    \caption{Maximum observed execution time vs. number of rows prepared by the \Jenga attacker}
    \label{fig:superloop}
\end{figure*}

\subsection{The \Jenga Attack - Superloop Setup Results}
\label{sec:results:superloop}

Figure~\ref{fig:superloop} presents the results obtained in the first experiment for \CR{6} TACLeBench workloads (4 \CR{additional graphs are available on the \href{https://gitlab.inria.fr/vabgrall/jenga}{\Jenga gitlab repository}}). These benchmarks were selected to cover different execution-time scales and memory-access behaviours. In particular, their baseline execution times \CR{span approximately two orders of magnitude}, allowing us to evaluate the impact of the proposed mechanism across short, medium, and longer-running workloads. Moreover, the workloads exhibit distinct memory-access patterns. \CR{For instance,} \textit{CountNegative} performs a tight sequential scan while \textit{LMS} and \textit{FFT} rely on repeated burst-like accesses over small groups of data. For \CR{all evaluated} workloads, as the number of rows whose PRAC counters are manipulated by the attacker increases, we observe that sufficiently large delays are induced to cause execution time (blue-dashed line) to exceed the pWCET (red-dotted line), which was computed with a confidence level of 0.9999. 
We also observe a plateau in the increase in execution time. This behaviour is explained by the bound derived in Equation~\ref{eq:majorant_2}: The workloads do not generate enough memory requests to trigger additional \ac{RFM} commands beyond this point.

Additionally, we compute the theoretical PRAC-\CR{and-attacker-budget-aware} \ac{WCET} bound (green-dotted line) using the formula derived in Section~\ref{sec:wcet_analysis}, with $t_{RFM}^{\CR{max}} = \CR{t_{RFM} + t_{RCD} = }350 \CR{+ 15}$ns, $t_{REF}^{max}$ \CR{=} $t_{REF}$ $\CR{+}$ $\CR{t_{RCD}} = 195 + 15$ns, and $t_{REFI}=3900$ns, as found in the standard~\cite{jedec_prac} and in Ramulator's code. \CR{The additional $t_{RCD}$ term accounts for the row reopening latency after a refresh operation. Since refreshes close the previously activated row, it must be reactivated before memory accesses to this row can resume.} For all evaluated workloads, the derived bound safely exceeds the maximum observed execution time under attack conditions. \CR{Moreover, the bound follows the same trend as the observed execution times across workloads, although it} remains pessimistic for \textit{LMS}, \CR{\textit{Bitcount}} and \textit{CountNegative}.

\subsection{Effect of Initial Memory State on TACLeBench Workloads}
\label{sec:results:initmemstate}

\subsubsection{Number of Alerts Triggered}
\label{sec:results:numalerts}

\begin{figure*}[t]
    \centering

    \begin{subfigure}[t]{0.49\textwidth}
        \centering
        \includegraphics[width=\linewidth]{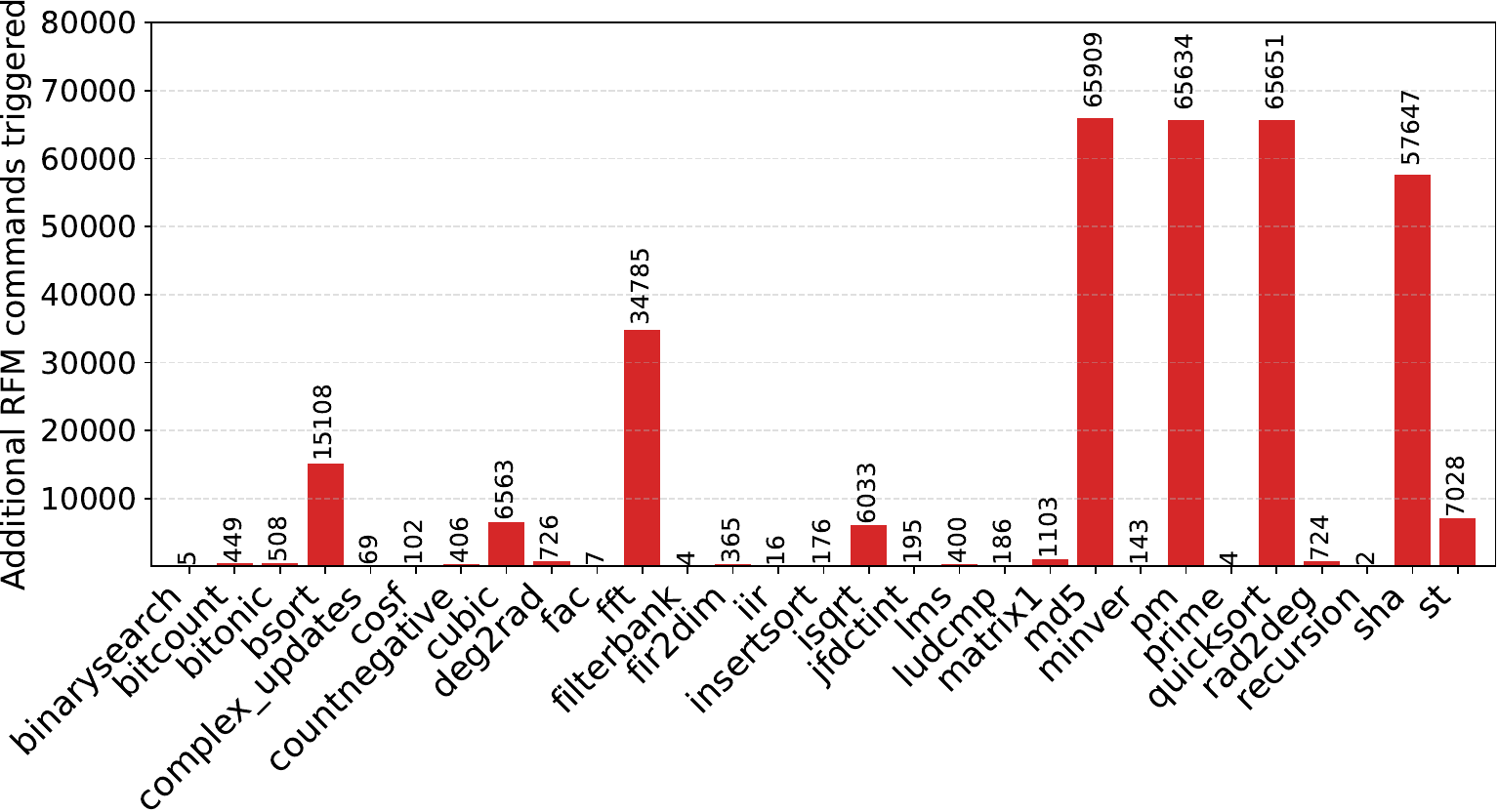}
        \caption{\footnotesize Kernel workloads}
    \end{subfigure}
    \hfill
    \begin{subfigure}[t]{0.49\textwidth}
        \centering
        \includegraphics[width=\linewidth]{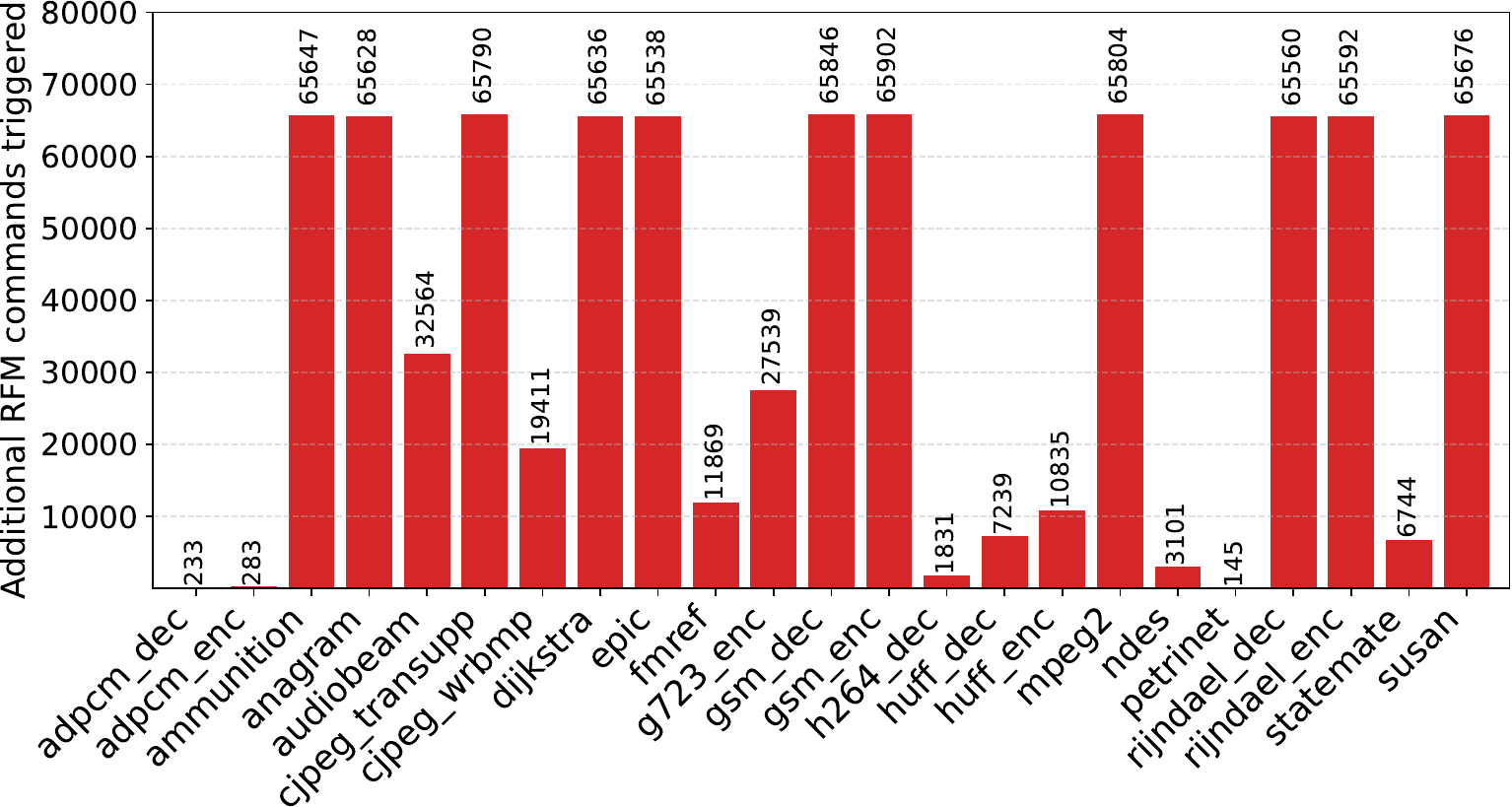}
        \caption{\footnotesize Sequential workloads}
    \end{subfigure}

    \caption{Number of additional alerts triggered under an adversarial initial memory state (with $T_{RH} = 128$).}
    \label{fig:PRAC_number}
    \vspace*{-1em}
\end{figure*}

\figurename~\ref{fig:PRAC_number} presents the number of additional alerts triggered when workloads are executed under an adversarial initial memory state instead of a clean initial state. As expected, the adversarial memory state systematically results in a higher number of alerts. This increase is proportional to the number of memory accesses performed by the executed workload.

Moreover, the number of additional alerts saturates to 65000 for some workloads, such as \textit{dijkstra} and \textit{anagram}. This value corresponds to the bound derived in Equation \ref{eq:majorant_1} with $n_{RFM} = 1$, $N = 1$, and $r_{bank} = 6\CR{5}536$. In these cases, the entire \ac{RFM} cascade was triggered, meaning that the workload generated enough memory accesses to fully exploit the adversarial memory state.

Conversely, workloads performing only a small number of memory accesses trigger significantly fewer additional alerts, which is consistent with the bound derived in Equation \ref{eq:majorant_2}. For example, the \textit{filterbank} workload performs only seven memory accesses and triggers 4 additional alerts under the adversarial initial memory state.

\subsubsection{Increase in Execution Time}
\label{sec:results:exectime}

\begin{figure}[t]
    \centering
        \includegraphics[width=0.95\columnwidth,trim={20pt 21pt 18pt 20pt},clip]{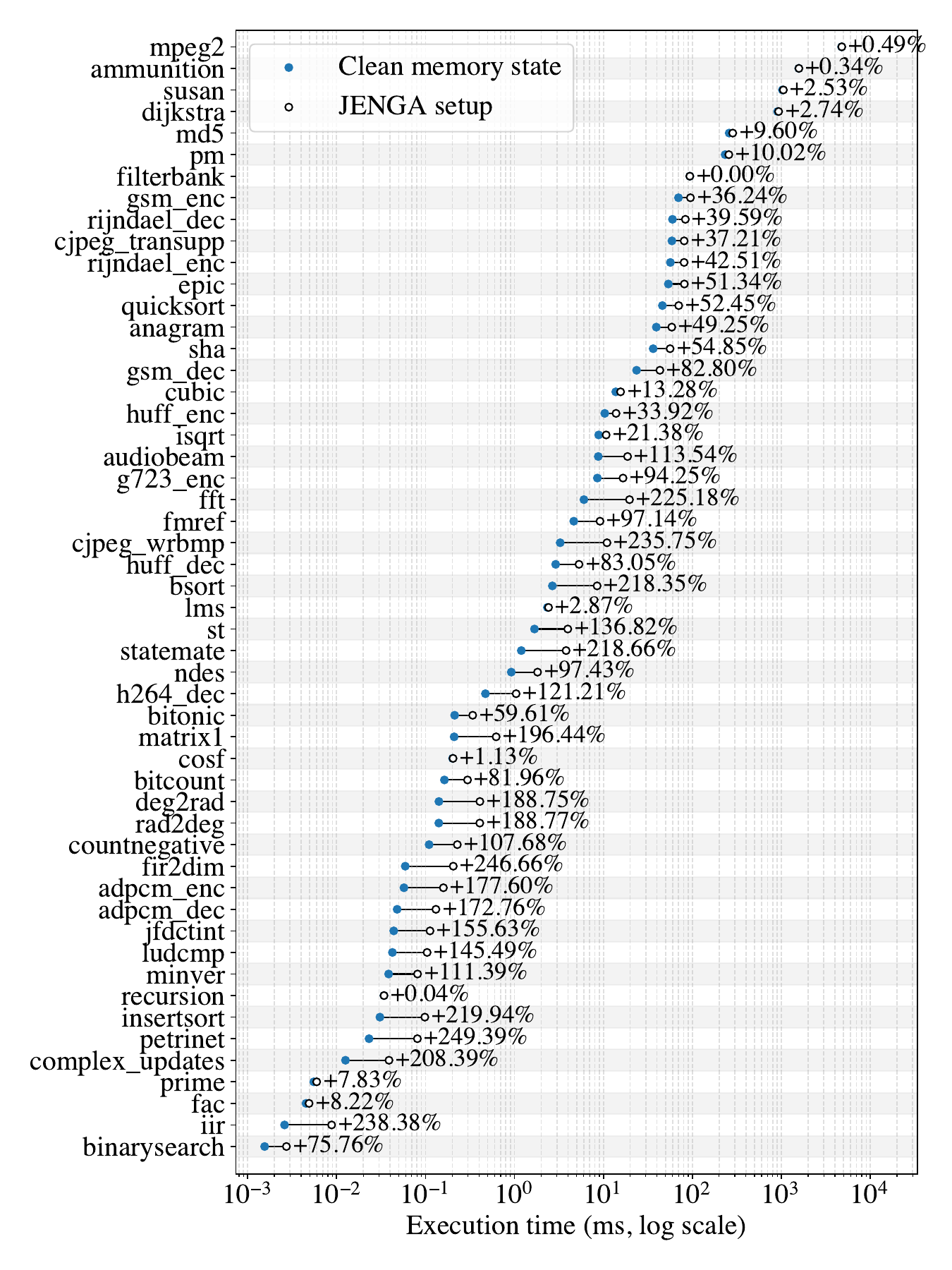}
    \caption{Execution time (log scale) for the baseline case with a clean memory and increase under an adversarial initial memory state.}
    \label{fig:exec_time}
    \vspace{-1em}
\end{figure}

\figurename~\ref{fig:exec_time} presents both the baseline execution time obtained with a clean memory state and the increase in execution time observed under an adversarial initial memory state (i.e., under the \Jenga attack). The impact of the adversarial memory state varies significantly depending on the workload executed. Across all evaluated benchmarks, the execution time increase ranges from nearly 0\% up to 249\% for the \textit{petrinet} workload.

Workloads exhibiting a negligible increase in execution time can be divided into two categories. The first category includes computation-intensive workloads that perform relatively few memory accesses, such as \textit{filterbank}. Since these workloads interact minimally with DRAM, they only trigger a few additional mitigations under the adversarial memory state. The second category includes workloads that already have a very large baseline execution time, even under a clean memory state. In such cases, although additional \ac{RFM} commands are triggered, the resulting overhead remains small relative to the total execution time. This behaviour can be observed for workloads such as \textit{ammunition} and \textit{dijkstra}. 

Conversely, for the majority of the evaluated workloads, the adversarial memory state induces substantial execution-time increases, demonstrating that mitigation-induced delays can significantly affect timing behaviour in simple real-time system configurations.

\subsubsection{Influence of $T_{RH}$}

\begin{figure*}[t]
    \centering
    \includegraphics[trim=0.5cm 0.85cm 0.61cm 0.62cm, clip, width=\linewidth]{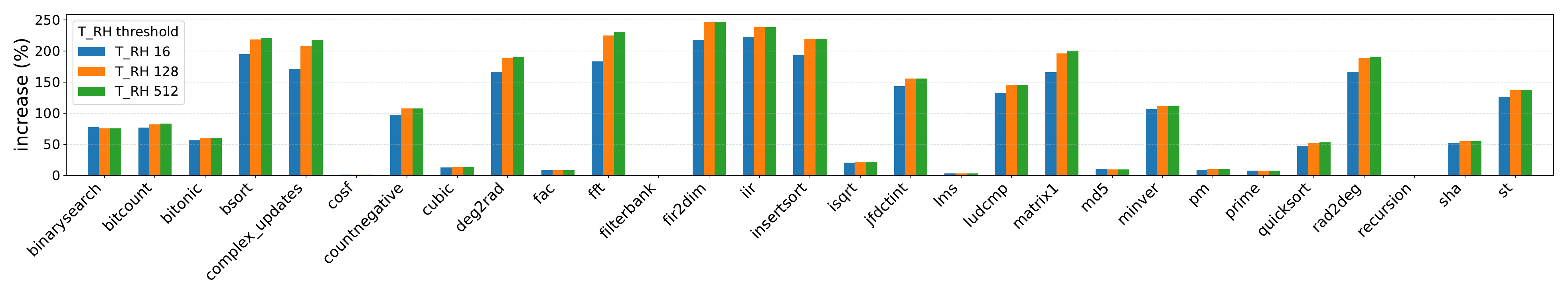}

    \caption{Increase in execution time under an adversarial initial memory state for kernel workloads and different values of $T_{RH}$.}
    \label{fig:trh_influence}
    \vspace*{-1em}
\end{figure*}

We evaluated the influence of $T_{RH}$ on the execution time difference between clean and adversarial initial memory states. The results are presented in \figurename~\ref{fig:trh_influence}. We observe that the impact of an adversarial memory state becomes significantly more pronounced for larger $T_{RH}$ values, i.e., when the DRAM is assumed to be less vulnerable to read-disturbance effects.
This behaviour can be explained by the fact that low $T_{RH}$ values cause workloads to trigger a large number of alerts even under a clean initial memory state. Consequently, the difference in the number of \ac{RFM} commands issued between clean and adversarial states remains relatively limited. Conversely, when $T_{RH}$ is high, workloads trigger far fewer mitigations under normal conditions, making the additional alerts induced by the adversarial memory state proportionally much more significant.
However, larger $T_{RH}$ values also make the construction of an adversarial memory state considerably more expensive for the attacker, as bringing PRAC counters close to the activation threshold requires a substantially higher number of memory accesses.

\section{Discussion}
\label{sec:discussion}

\subsection{Feasibility of the Attack}

Several factors could make the practical implementation of the \Jenga attack challenging. The first major challenge lies in the interaction between periodic REF commands and the PRAC counters. Indeed, periodic refreshes increment PRAC counters, thereby perturbing the memory state manipulated by the attacker. \CR{However, natural refreshes are issued periodically at fixed $t_{REFI}$ intervals, and every row is refreshed once during each $t_{REFW}$ window. Consequently, after determining the initial state of the relevant PRAC counters, an attacker can keep track of their evolution over time by monitoring the elapsed number of $t_{REFW}$ windows. Determining this initial state constitutes a second challenge, since PRAC counters are internal DRAM structures that are not directly accessible from software, including the operating system. Nevertheless, direct access to the counters is not required by the attack, which only assumes user-level privileges. Instead, we believe the initial counter state could be inferred by exploiting the variations in memory access latency induced by mitigative refresh operations. By repeatedly accessing selected rows and monitoring memory access latency, an attacker can detect when mitigative refreshes are triggered and, consequently, when the corresponding PRAC counters are reset. Previous work has already demonstrated that such mitigative refresh operations are observable from user space and can be exploited to construct side and covert channels~\cite{bostanci_UnderstandingMitigating_CCSCA_2025}. Although mechanisms such as FR-RFM, randomized counter resets, or bank-local preventive actions may complicate this timing side channel~\cite{bostanci_UnderstandingMitigating_CCSCA_2025}, they do not eliminate the timing interference introduced by mitigative refreshes. Therefore, while these mechanisms may increase the practical complexity of the attack, they do not fundamentally invalidate the assumptions underlying \Jenga.}


Another challenge comes from the internal subarray organisation of DRAM banks. In practice, a DRAM bank is divided into multiple subarrays \cite{subarray}, and neighbouring rows located in different subarrays may not interact in the same way as rows within a single subarray. Consequently, an \ac{RFM} command targeting the last row of a subarray may not refresh (and therefore activate) the first row of the following subarray. In such cases, the propagation of an \ac{RFM} cascade would stop at subarray boundaries, limiting the maximum size of the "tower" that an attacker can construct. \CR{This limitation can be accounted for in the \textit{attacker's budget} $r_{attacker}$, which represents the maximum number of rows that can effectively be prepared by the attacker. In particular, $r_{attacker}$ can be lower than the total number of rows available in a DRAM bank due to subarray boundaries.} Nevertheless, this limitation does not completely prevent the attack. Modern DRAM subarrays typically contain on the order of several hundreds to nearly one thousand rows \cite{subarray}, which still leaves room for sufficiently large \ac{RFM} cascades capable of introducing non-negligible timing overheads. Moreover, an attacker could potentially prepare multiple cascades across different subarrays of the same bank, thereby \CR{better exploiting the available attacker budget and} partially compensating for the interruption of propagation at subarray boundaries.

Finally, the attacker must know the physical-to-DRAM-layout mapping in order to construct the \Jenga tower. Such mappings are typically proprietary and undocumented. However, prior work has shown that it is possible to reverse-engineer real-world DRAM mappings by exploiting row-buffer conflict side channels \cite{knock_knock,drama}.

Overall, the attack remains feasible in practice, and the statefulness of counter-based RowHammer countermeasures can introduce significant timing variations in real-time systems. Our results provide valuable insight into the interaction between RowHammer countermeasures and timing predictability, opening new opportunities for the design of more predictable real-time systems.

\subsection{\CR{Mitigative Refresh Granularity} and Memory Throttling}

The attacker does not need to share the same physical DRAM region as the victim task. The attacker can construct the \Jenga tower inside its own allocated memory region and trigger its collapse before the victim task executes. \CR{In our system model, we assume that, upon receiving an \ac{ABO} alert, the memory controller issues all-bank $RFM_{ab}$ commands, which stall the entire memory subsystem. This assumption is consistent with the DDR5 JEDEC standard \cite{jedec_prac}, where \ac{ABO} alerts do not identify the bank containing the aggressor rows. Nevertheless, if future revisions of the standard or proprietary implementations were to identify the aggressor bank in the \ac{ABO} alert and issue same-bank $RFM_{sb}$ commands instead, the attack would remain possible. Same-bank $RFM_{sb}$ commands target all banks sharing the same bank index across different bank groups. Therefore, the attacker would need to share the same bank index as the victim for the attack to be successful, even if the two tasks reside in different bank groups.}

Finally, some systems employ memory-throttling mechanisms to bound the number of memory accesses that a task can perform in a given time window. MemGuard \cite{memguard}, for instance, enforces such quotas in multi-core systems. These mechanisms reduce the rate at which an attacker can prepare the adversarial memory state, but do not fundamentally prevent the attack. Instead, the attacker could progressively manipulate different rows across multiple time windows until a sufficiently adverse state is reached. Successfully carrying out such an attack would, however, require tracking or predicting the evolution of the memory state while the attacker is not executing, including the effects of genuine tasks and periodic REF commands. Investigating the feasibility of such long-term manipulations constitutes an interesting direction for future work.

\subsection{\CR{Complex Architectures: Multi-Core Systems and Memory Hierarchies}}

In this work, we focused on a single-core system without caches \CR{and without an operating system} in order to isolate and characterise the impact of counter-based RowHammer countermeasures on execution time. While such assumptions remain relevant for some critical embedded systems prioritising predictability, many modern real-time platforms rely on multi-core architectures, \CR{(real-time) operating systems,} and cache hierarchies. \CR{Nevertheless, these features do not fundamentally invalidate the attack.} 

\CR{In multi-core systems, concurrent accesses from other cores may delay the attacker's memory accesses and introduce additional timing variability. However, the attack does not require precise timing control: the attacker only needs to drive the PRAC counters to specific values before triggering the cascade, determine their initial counter state, and account for periodic refreshes affecting the attacker's memory region. Furthermore, memory partitioning and isolation mechanisms commonly used in safety-critical systems can allow an attacker to construct \Jenga towers within its allocated memory region without interference from other processes.}

In systems with caches, a practical attacker would additionally need to ensure that memory accesses effectively reach DRAM rather than being absorbed by the cache hierarchy. This could potentially be achieved through techniques such as explicit cache flushing instructions, cache-eviction patterns, or uncached memory accesses, depending on the capabilities exposed by the target platform \cite{one_loc_rh}.

\CR{Although we expect the \Jenga attack to remain applicable on more complex systems, evaluating its impact under these architectures is an important direction for future work. Our current WCET analysis focuses on a simple architecture and does not account for the effects introduced by more complex processor and memory hierarchies. For instance, cache hits may mask part of the RFM-induced latency overhead, while pipeline overlap and out-of-order execution may reduce the observable impact of DRAM stalls. Extending the analysis to account for these mechanisms would enable tighter and less pessimistic WCET bounds.}

\section{Conclusion} 
\label{sec:conclusion} 

In this paper, we studied the impact of counter-based RowHammer mitigations on the timing behaviour of safety-critical real-time systems. While previous works on RowHammer mainly focused on security, we showed that preventive mitigations themselves can introduce significant timing overheads that must be accounted for in \ac{WCET} analysis. 

To demonstrate this effect, we introduced the \Jenga attack, in which an attacker-controlled task progressively manipulates the PRAC counter\CR{s} state in order to maximise the execution time of a genuine real-time task. Our experimental evaluation on TACLeBench \TR{workloads} showed that the initial memory state can have a substantial impact on execution time, with workloads experiencing overheads exceeding up to 200\% under adversarial conditions. \TR{We also showed that the impact of adversarial memory states becomes more pronounced as the RowHammer detection threshold ($T_{RH}$) increases, despite the increased effort required from the attacker to construct such states.} 
We also proposed a method to integrate deterministic counter-based countermeasures into the computation of a program's \ac{WCET} when executing on DRAM-protected systems. We then applied this methodology to the PRAC-N mitigation mechanism \CR{standardised for} DDR5 memories and derived analytical bounds on the number of mitigative refreshes that may be triggered during program execution. 

\TR{Our results show that the statefulness of counter-based RowHammer mitigations can create timing variations that challenge the predictability requirements of safety-critical real-time systems.} More generally, this work highlights the importance of jointly considering security mechanisms and timing guarantees, as the protections  themselves may introduce new timing-related attack surfaces. 

Future work will focus on extending the analysis to more complex platforms, including systems with \CR{an OS,} caches and multi-core architectures, where additional sources of contention and interference may further amplify or mitigate the observed effects. \TR{We also plan to investigate the practical feasibility of the \Jenga attack on real DDR5 devices to validate our observations beyond simulation and better characterise the impact of DRAM implementation details on the attack behaviour.}

\section*{Acknowledgments}

We thank RTSS 2026 anonymous reviewers for their insightful comments and suggestions, which helped improve the clarity and presentation of this work.

\FloatBarrier
\bibliographystyle{IEEEtranNoDash}
\bibliography{sample}

\end{document}